\documentclass[twoside]{article}
\font\msbm=msbm10 scaled 1000      
\font\eurm=eurm10 scaled 1000 	   
\def\Re{\mbox{\rm Re}\, }
\def\scriptRe{\mbox{\rm\scriptsize Re}\,}
\def\Im{\mbox{\rm Im}\, }
\def\sn{\mbox{\rm sn}\, }
\def\varkappa{\mbox{\msbm\char'173}}

\newcommand{\Bbb}[1]{\mbox{\msbm #1}\,}
\newcommand{\Eu}[1]{\mbox{\eurm #1}}

\newtheorem{corollary}{Corollary}[section]
\newtheorem{theorem}{Theorem}[section]

\newtheorem{@remark}{\bf Remark}[section]

\def\dsp{\displaystyle}
\def\scr{\scriptstyle}

\def\longto{\mathop{\longrightarrow}}
\def\HAT#1{\hskip-1.2em\hbox to1.3em{\vbox{\msbm\char'134\vskip3pt}\hss}#1}
\def\Pso{\displaystyle\mathop{\Psi}}
\def\Theto{\displaystyle\mathop{\Theta}}

\def\temp{1.31}
\let\tempp=\relax
\expandafter\ifx\csname psboxversion\endcsname\relax
\else
    \ifdim\temp cm>\psboxversion cm
    \else
      \let\temp=\psboxversion
      \let\tempp= 
    \fi
\fi
\tempp
\let\psboxversion=\temp
\catcode`\@=11
\def\execute#1{#1}
\def\psm@keother#1{\catcode`#112\relax}
\def\executeinspecs#1{%
\execute{\begingroup\let\do\psm@keother\dospecials\catcode`\^^M=9#1\endgroup}}
\def\psfortextures{
\def\PSspeci@l##1##2{%
\special{illustration ##1\space scaled ##2}%
}}
\def\psfordvitops{
\def\PSspeci@l##1##2{%
\special{dvitops: import ##1\space \the\drawingwd \the\drawinght}%
}}
\def\psfordvips{
\def\PSspeci@l##1##2{%
\d@my=0.1bp \d@mx=\drawingwd \divide\d@mx by\d@my%
\includegraphics{##1\space}%
}}
\def\psforoztex{
\def\PSspeci@l##1##2{%
\special{##1 \space
      ##2 1000 div dup scale
      \putsp@ce{\number-\psllx} \putsp@ce{\number-\pslly} translate
}%
}}
\def\putsp@ce#1{#1 }
\def\psfordvitps{
\def\psdimt@n@sp##1{\d@mx=##1\relax\edef\psn@sp{\number\d@mx}}
\def\PSspeci@l##1##2{%
\special{dvitps: Include0 "psfig.psr"}
\psdimt@n@sp{\drawingwd}
\special{dvitps: Literal "\psn@sp\space"}
\psdimt@n@sp{\drawinght}
\special{dvitps: Literal "\psn@sp\space"}
\psdimt@n@sp{\psllx bp}
\special{dvitps: Literal "\psn@sp\space"}
\psdimt@n@sp{\pslly bp}
\special{dvitps: Literal "\psn@sp\space"}
\psdimt@n@sp{\psurx bp}
\special{dvitps: Literal "\psn@sp\space"}
\psdimt@n@sp{\psury bp}
\special{dvitps: Literal "\psn@sp\space startTexFig\space"}
\special{dvitps: Include1 "##1"}
\special{dvitps: Literal "endTexFig\space"}
}}
\def\psforDVIALW{
\def\PSspeci@l##1##2{
\special{language "PS"
literal "##2 1000 div dup scale"
include "##1"}}}
\def\psonlyboxes{
\def\PSspeci@l##1##2{%
\at(0cm;0cm){\boxit{\vbox to\drawinght
  {\vss
  \hbox to\drawingwd{\at(0cm;0cm){\hbox{(##1)}}\hss}
  }}}
}%
}
\def\psloc@lerr#1{%
\let\savedPSspeci@l=\PSspeci@l%
\def\PSspeci@l##1##2{%
\at(0cm;0cm){\boxit{\vbox to\drawinght
  {\vss
  \hbox to\drawingwd{\at(0cm;0cm){\hbox{(##1) #1}}\hss}
  }}}
\let\PSspeci@l=\savedPSspeci@l
}%
}
\newread\pst@mpin
\newdimen\drawinght\newdimen\drawingwd
\newdimen\psxoffset\newdimen\psyoffset
\newbox\drawingBox
\newif\ifNotB@undingBox
\newhelp\PShelp{Proceed: you'll have a 5cm square blank box instead of
your graphics (Jean Orloff).}
\def\@mpty{}
\def\s@tsize#1 #2 #3 #4\@ndsize{
  \def\psllx{#1}\def\pslly{#2}%
  \def\psurx{#3}\def\psury{#4}
  \ifx\psurx\@mpty\NotB@undingBoxtrue
  \else
    \drawinght=#4bp\advance\drawinght by-#2bp
    \drawingwd=#3bp\advance\drawingwd by-#1bp
  \fi
  }
\def\sc@nline#1:#2\@ndline{\edef\p@rameter{#1}\edef\v@lue{#2}}
\def\g@bblefirstblank#1#2:{\ifx#1 \else#1\fi#2}
\def\psm@keother#1{\catcode`#112\relax}
\def\execute#1{#1}
{\catcode`\%=12
\xdef\B@undingBox{
}   
\def\ReadPSize#1{
 \edef\PSfilename{#1}
 \openin\pst@mpin=#1\relax
 \ifeof\pst@mpin \errhelp=\PShelp
   \errmessage{I haven't found your postscript file (\PSfilename)}
   \psloc@lerr{was not found}
   \s@tsize 0 0 142 142\@ndsize
   \closein\pst@mpin
 \else
   \immediate\write\psbj@inaux{#1,}
   \loop
     \executeinspecs{\catcode`\ =10\global\read\pst@mpin to\n@xtline}
     \ifeof\pst@mpin
       \errhelp=\PShelp
       \errmessage{(\PSfilename) is not an Encapsulated PostScript File:
           I could not find any \B@undingBox: line.}
       \edef\v@lue{0 0 142 142:}
       \psloc@lerr{is not an EPSFile}
       \NotB@undingBoxfalse
     \else
       \expandafter\sc@nline\n@xtline:\@ndline
       \ifx\p@rameter\B@undingBox\NotB@undingBoxfalse
         \edef\t@mp{%
           \expandafter\g@bblefirstblank\v@lue\space\space\space}
         \expandafter\s@tsize\t@mp\@ndsize
       \else\NotB@undingBoxtrue
       \fi
     \fi
   \ifNotB@undingBox\repeat
   \closein\pst@mpin
 \fi
\message{#1}
}
\newcount\xscale \newcount\yscale \newdimen\pscm\pscm=1cm
\newdimen\d@mx \newdimen\d@my
\let\ps@nnotation=\relax
\def\psboxto(#1;#2)#3{\vbox{
   \ReadPSize{#3}
   \divide\drawingwd by 1000
   \divide\drawinght by 1000
   \d@mx=#1
   \ifdim\d@mx=0pt\xscale=1000
         \else \xscale=\d@mx \divide \xscale by \drawingwd\fi
   \d@my=#2
   \ifdim\d@my=0pt\yscale=1000
         \else \yscale=\d@my \divide \yscale by \drawinght\fi
   \ifnum\yscale=1000
         \else\ifnum\xscale=1000\xscale=\yscale
                    \else\ifnum\yscale<\xscale\xscale=\yscale\fi
              \fi
   \fi
   \divide \psxoffset by 1000\multiply\psxoffset by \xscale
   \divide \psyoffset by 1000\multiply\psyoffset by \xscale
   \global\divide\pscm by 1000
   \global\multiply\pscm by\xscale
   \multiply\drawingwd by\xscale \multiply\drawinght by\xscale
   \ifdim\d@mx=0pt\d@mx=\drawingwd\fi
   \ifdim\d@my=0pt\d@my=\drawinght\fi
   \message{scaled \the\xscale}
 \hbox to\d@mx{\hss\vbox to\d@my{\vss
   \global\setbox\drawingBox=\hbox to 0pt{\kern\psxoffset\vbox to 0pt{
      \kern-\psyoffset
      \PSspeci@l{\PSfilename}{\the\xscale}
      \vss}\hss\ps@nnotation}
   \global\ht\drawingBox=\the\drawinght
   \global\wd\drawingBox=\the\drawingwd
   \baselineskip=0pt
   \copy\drawingBox
 \vss}\hss}
  \global\psxoffset=0pt
  \global\psyoffset=0pt
  \global\pscm=1cm
  \global\drawingwd=\drawingwd
  \global\drawinght=\drawinght
}}
\def\psboxscaled#1#2{\vbox{
  \ReadPSize{#2}
  \xscale=#1
  \message{scaled \the\xscale}
  \divide\drawingwd by 1000\multiply\drawingwd by\xscale
  \divide\drawinght by 1000\multiply\drawinght by\xscale
  \divide \psxoffset by 1000\multiply\psxoffset by \xscale
  \divide \psyoffset by 1000\multiply\psyoffset by \xscale
  \global\divide\pscm by 1000
  \global\multiply\pscm by\xscale
  \global\setbox\drawingBox=\hbox to 0pt{\kern\psxoffset\vbox to 0pt{
     \kern-\psyoffset
     \PSspeci@l{\PSfilename}{\the\xscale}
     \vss}\hss\ps@nnotation}
  \global\ht\drawingBox=\the\drawinght
  \global\wd\drawingBox=\the\drawingwd
  \baselineskip=0pt
  \copy\drawingBox
  \global\psxoffset=0pt
  \global\psyoffset=0pt
  \global\pscm=1cm
  \global\drawingwd=\drawingwd
  \global\drawinght=\drawinght
}}
\def\psbox#1{\psboxscaled{1000}{#1}}
\newif\ifn@teof\n@teoftrue
\newif\ifc@ntrolline
\newif\ifmatch
\newread\j@insplitin
\newwrite\j@insplitout
\newwrite\psbj@inaux
\immediate\openout\psbj@inaux=psbjoin.aux
\immediate\write\psbj@inaux{\string\joinfiles}
\immediate\write\psbj@inaux{\jobname,}
\immediate\let\oldinput=\input
\def\input#1 {
 \immediate\write\psbj@inaux{#1,}
 \oldinput #1 }
\def\empty{}
\def\setmatchif#1\contains#2{
  \def\match##1#2##2\endmatch{
    \def\tmp{##2}
    \ifx\empty\tmp
      \matchfalse
    \else
      \matchtrue
    \fi}
  \match#1#2\endmatch}
\def\warnopenout#1#2{
 \setmatchif{TrashMe,psbjoin.aux,psbjoin.all}\contains{#2}
 \ifmatch
 \else
   \immediate\openin\pst@mpin=#2
   \ifeof\pst@mpin
     \else
     \errhelp{If the content of this file is so precious to you, abort (ie
press x or e) and rename it before retrying.}
     \errmessage{I'm just about to replace your file named #2}
   \fi
   \immediate\closein\pst@mpin
 \fi
 \message{#2}
 \immediate\openout#1=#2}
{
\catcode`\%=12
\gdef\splitfile#1 {
 \immediate\openin\j@insplitin=#1
 \message{Splitting file #1 into:}
 \warnopenout\j@insplitout{TrashMe}
 \loop
   \ifeof
     \j@insplitin\immediate\closein\j@insplitin\n@teoffalse
   \else
     \n@teoftrue
     \executeinspecs{\global\read\j@insplitin to\spl@tinline\expandafter
       \ch@ckbeginnewfile\spl@tinline
     \ifc@ntrolline
     \else
       \toks0=\expandafter{\spl@tinline}
       \immediate\write\j@insplitout{\the\toks0}
     \fi
   \fi
 \ifn@teof\repeat
 \immediate\closeout\j@insplitout}
\gdef\ch@ckbeginnewfile#1
 \def\t@mp{#1}
 \ifx\empty\t@mp
   \def\t@mp{#3}
   \ifx\empty\t@mp
     \global\c@ntrollinefalse
   \else
     \immediate\closeout\j@insplitout
     \warnopenout\j@insplitout{#2}
     \global\c@ntrollinetrue
   \fi
 \else
   \global\c@ntrollinefalse
 \fi}
\gdef\joinfiles#1\into#2 {
 \message{Joining following files into}
 \warnopenout\j@insplitout{#2}
 \message{:}
 {
 \edef\w@##1{\immediate\write\j@insplitout{##1}}
 \w@{
 \w@{
 \w@{
 \w@{
 \w@{
 \w@{
 \w@{
 \w@{
 \w@{\string\input\space psbox.tex}
 \w@{\string\splitfile{\string\jobname}}
 }
 \tre@tfilelist#1, \endtre@t
 \immediate\closeout\j@insplitout}
\gdef\tre@tfilelist#1, #2\endtre@t{
 \def\t@mp{#1}
 \ifx\empty\t@mp
   \else
   \llj@in{#1}
   \tre@tfilelist#2, \endtre@t
 \fi}
\gdef\llj@in#1{
 \immediate\openin\j@insplitin=#1
 \ifeof\j@insplitin
   \errmessage{I couldn't find file #1.}
   \else
   \message{#1}
   \toks0={
   \immediate\write\j@insplitout{\the\toks0}
   \executeinspecs{\global\read\j@insplitin to\oldj@ininline}
   \loop
     \ifeof\j@insplitin\immediate\closein\j@insplitin\n@teoffalse
       \else\n@teoftrue
       \executeinspecs{\global\read\j@insplitin to\j@ininline}
       \toks0=\expandafter{\oldj@ininline}
       \let\oldj@ininline=\j@ininline
       \immediate\write\j@insplitout{\the\toks0}
     \fi
   \ifn@teof
   \repeat
   \immediate\closein\j@insplitin
 \fi}
}
\def\autojoin{
 \immediate\write\psbj@inaux{\string\into\space psbjoin.all}
 \immediate\closeout\psbj@inaux
 \input psbjoin.aux
}
\def\centinsert#1{\midinsert\line{\hss#1\hss}\endinsert}
\def\psannotate#1#2{\def\ps@nnotation{#2\global\let\ps@nnotation=\relax}#1}
\def\pscaption#1#2{\vbox{
   \setbox\drawingBox=#1
   \copy\drawingBox
   \vskip\baselineskip
   \vbox{\hsize=\wd\drawingBox\setbox0=\hbox{#2}
     \ifdim\wd0>\hsize
       \noindent\unhbox0\tolerance=5000
    \else\centerline{\box0}
    \fi
}}}
\def\psfig#1#2#3{\pscaption{\psannotate{#1}{#2}}{#3}}
\def\psfigurebox#1#2#3{\pscaption{\psannotate{\psbox{#1}}{#2}}{#3}}
\def\at(#1;#2)#3{\setbox0=\hbox{#3}\ht0=0pt\dp0=0pt
  \rlap{\kern#1\vbox to0pt{\kern-#2\box0\vss}}}
\newdimen\gridht \newdimen\gridwd
\def\gridfill(#1;#2){
  \setbox0=\hbox to 1\pscm
  {\vrule height1\pscm width.4pt\leaders\hrule\hfill}
  \gridht=#1
  \divide\gridht by \ht0
  \multiply\gridht by \ht0
  \gridwd=#2
  \divide\gridwd by \wd0
  \multiply\gridwd by \wd0
  \advance \gridwd by \wd0
  \vbox to \gridht{\leaders\hbox to\gridwd{\leaders\box0\hfill}\vfill}}
\def\fillinggrid{\at(0cm;0cm){\vbox{
  \gridfill(\drawinght;\drawingwd)}}}
\def\textleftof#1:{
  \setbox1=#1
  \setbox0=\vbox\bgroup
    \advance\hsize by -\wd1 \advance\hsize by -2em}
\def\textrightof#1:{
  \setbox0=#1
  \setbox1=\vbox\bgroup
    \advance\hsize by -\wd0 \advance\hsize by -2em}
\def\endtext{
  \egroup
  \hbox to \hsize{\valign{\vfil##\vfil\cr%
\box0\cr%
\noalign{\hss}\box1\cr}}}
\def\frameit#1#2#3{\hbox{\vrule width#1\vbox{
  \hrule height#1\vskip#2\hbox{\hskip#2\vbox{#3}\hskip#2}%
        \vskip#2\hrule height#1}\vrule width#1}}
\def\boxit#1{\frameit{0.4pt}{0pt}{#1}}
\catcode`\@=12 
 \psfordvips   

\def\theequation{\arabic{section}.\arabic{equation}}

\begin{document}
\pagenumbering{arabic}
\thispagestyle{empty}

\centerline{\Large The nonlinear steepest descent approach} 
\centerline{\Large to the asymptotics of the second Painlev\'e}
\centerline{\Large transcendent in the complex domain}
\vskip.2in
\centerline{A.~Its}
\centerline{{\it Department of Mathematical Sciences,}}
\centerline{{\it Indiana University -- Purdue University  Indianapolis}}
\centerline{{\it Indianapolis, IN 46202-3216, USA}}
\centerline{A.~Kapaev}
\centerline{{\it St.~Petersburg Branch of Steklov Mathematical Institute,}}
\centerline{{\it Russian Academy of Sciences,}}
\centerline{{\it St. Petersburg, 191011, Russia}}

\begin{abstract}
The asymptotics of the generic second Painlev\'e transcendent $u(x)$ as 
$|x|\to\infty$, $\arg x\in\bigl(\frac{\pi}{3}k,\frac{\pi}{3}(k+1)\bigr)$, 
$k=0,1,\dots,5$ is found and justified via the direct asymptotic analysis 
of the associated Riemann-Hilbert problem based on the Deift-Zhou nonlinear 
steepest descent method. The asymptotics is proved of the Boutroux type,
i.e.\ it is expressed in terms of the elliptic functions. 
Kapaev-Novokshenov's explicit connection formulae between the asymptotic 
phases in the different sectors are obtained as well.
\end{abstract}

\section{Introduction}
\label{intro}
\setcounter{equation}{0}

The large argument asymptotic behavior of the Painlev\'e functions of the 
first and second kinds were first described in the classical work of 
P.~Bou\-tro\-ux \cite{boutroux}. In particular, Boutroux found a general 
2-parameter elliptic (i.e.\ expressed in terms of elliptic functions) 
asymptotic formulae for the Painlev\'e functions. 

In 1988, using the multiscale expansion method, N.~Joshi and M.~Kruskal 
\cite{joshi_kruskal} revised the result of Boutroux and found the phase shift 
in the Boutroux elliptic ansatz. Specifically, Joshi and Kruskal showed that 
in the case of the second Painlev\'e equation,
\begin{equation}\label{p2}
\frac{d^2u}{dx^2}=2u^3+xu,
\end{equation}
the large $|x|$ asymptotics of its generic solution $u(x)$ is given by the 
following equation, 
\begin{equation}\label{bjk}
u(x)\sim c_{1}\sqrt{x}\,
\sn\bigl(\frac{2}{3}c_{2}x^{3/2}+\mu\omega_1+\nu\omega_2|D\bigr).
\end{equation} 
Here the periods $\omega_j$, $j=1,2$ of the elliptic sine, the multipliers 
$c_{1,2}$, and the module $D$ are (transcendental) functions of $\arg x$, 
while the parameters $\mu$, $\nu$ are constant in the interior of any sector 
$\arg x\in\bigl(\frac{\pi}{3}k,\frac{\pi}{3}(k+1)\bigr)$, $k=0,1,\dots,5$.

The connection formulae between the constants $\mu$, $\nu$ in the ansatz 
(\ref{bjk}) corresponding to the different sectors were found by 
V.~Novokshenov \cite{nov1} and A.~Kapaev \cite{kapaev1, kapaev2} with the 
help of the isomonodromy deformation method which was introduced in the very 
beginning of the 80-s by H.~Flasch\-ka and A.~Newell \cite{fn} and by 
M.~Jimbo, T.~Miwa and K.~Ueno \cite{jmu, jm2, jm3}.

The isomonodromy deformation method associates with a given Painlev\'e 
equation an auxiliary system of linear ODEs with rational coefficients whose 
monodromy data are the first integrals of the Painlev\'e equation. Hence the 
asymptotic evaluation of the solutions of the Painlev\'e equations reduces to 
the asymptotic solution either of the direct or of the inverse monodromy 
problems for the auxiliary linear system. The relevant asymptotic scheme 
based on the asymptotic analysis of the direct monodromy problem was 
suggested in \cite{its_nov} and was used there and in the subsequent works of 
several other authors for the evaluation of the connection formulae for the 
Painlev\'e transcendent. We refer the reader to review \cite{its} for more 
on the history and apparatus of the isomonodromy method.

The approach of \cite{nov1} and \cite{kapaev1, kapaev2} is based on an 
extension of the direct monodromy problem technique of \cite{its_nov} to the 
asymptotic analysis of the Painlev\'e functions in the complex domain. The 
further development of this scheme was made in the works of A.~Kitaev 
\cite{kitaev} and A.~Kapaev and A.~Kitaev \cite{kap_kit} which were also 
devoted to the elliptic asymptotics of the Painlev\'e transcendents.

In the beginning of the 90s, P.~Deift and X.~Zhou proposed a nonlinear 
version of the classical steepest descent method for oscillatory 
Riemann-Hilbert problems \cite{DZ}. This allowed to develop an alternative 
to \cite{its_nov} asymptotic approach for solving the Painlev\'e equations 
based on the asymptotic solution of the relevant {\it inverse} monodromy 
problems, which can be formulated as the oscillatory matrix Riemann-Hilbert
(RH) problems. This approach has the advantage of not using any prior 
information of the solutions of the Painlev\'e equations, and it has already 
been applied to the trigonometric and exponentially decreasing asymptotics of 
the Painlev\'e functions on the real axes \cite{DZ2, IFK, its_kap}.
 
\medskip

The goal of this paper is to extend the Deift-Zhou method, in the case of the 
second Painlev\'e equation (\ref{p2}), to the complex domain. As a result we 
will obtain, rigorously and without any prior assumption crucial for 
\cite{nov1} and \cite{kapaev1, kapaev2}, the Boutroux type elliptic 
asymptotics together with the explicit formulae for the relevant phase shifts 
directly from the associated Riemann-Hilbert problem. 

The plan of the paper is as follows. In Section~\ref{RH_problem}, we remind 
the setting the RH problem related to equation (\ref{p2}) and formulate our 
main theorem (Theorem~\ref{theorem1}). In Section~\ref{RHp_transform}, we 
transform the RH problem to the one posed on the graph consisting of the 
anti-Stokes lines of a certain function $g(z)$ emanating from its critical 
points and connected by three finite Stokes lines. In 
Section~\ref{g_function}, we explicitly construct the function $g(z)$ 
mentioned{\footnote{In the most recent applications of the nonlinear steepest 
descent method to the orthogonal polynomials (see e.g.\ \cite{deift}) this 
step corresponds to the construction of the so called equilibrium measure}}. 
In Section~\ref{BA_function}, we construct the solution of the reduced RH 
problem on the finite Stokes lines of $g(z)$ in terms of the Jacobi theta 
functions. In Section~\ref{local_solutions}, we construct the approximate 
solutions of the RH problem near the critical points of $g(z)$ in terms of 
the Airy functions. In Section~\ref{app_RH_solution}, we collect all the 
functions together and prove that this combined function approximates the 
solution of the original RH problem. In Section~\ref{as_P2_solution}, we 
extract from the approximate solution of the RH problem the approximate 
solution of the second Painlev\'e equation.

\medskip

The authors dedicate this paper to Barry McCoy on the occasion of his 60th 
birthday. We are especially happy to do this since it was a remarkable 1977 
paper of McCoy, Tracy, and Wu \cite{mtw} where it was demonstrated for the 
very first time that an explicit and complete connection formulae for a 
Painlev\'e transcendent are possible.

\section{RH parametrization of the Painlev\'e functions of the second kind. 
Formulation of the main theorem}
\label{RH_problem}
\setcounter{equation}{0}

According to Flaschka and Newell \cite{fn} (see also \cite{fok_abl} and 
\cite{its_nov}) the Riemann-Hilbert problem associated with the second 
Painlev\'e equation (\ref{p2}) consists of finding the piecewise holomorphic 
$2\times 2$ matrix-valued function $\Psi(\lambda)$ of the complex variable 
$\lambda$ such that 
\begin{description}
\item{1)} $\Psi(\lambda)$ is holomorphic for 
$\lambda\in{\Bbb C}\setminus\cup\{\gamma_{k}\}$, where $\gamma_{k}$ are the 
rays $\gamma_k=\{\lambda\in{\Bbb C}\colon 
\arg\lambda=\frac{\pi}{6}+\frac{\pi}{3}(k-1)\}$, $k=1,\dots,6$, oriented from 
zero to infinity, and 
\begin{equation}\label{p21}
\Psi(\lambda)e^{\theta(\lambda,x)\sigma_3}\to I,\quad
\lambda\to\infty,\quad
\theta(\lambda,x)=i\bigl(\frac{4}{3}\lambda^3+x\lambda\bigr),
\end{equation}
$$
\sigma_3=\pmatrix{1&0\cr 0& -1\cr},
$$
\item{2)} on the rays $\gamma_k$ the jump conditions hold
\begin{equation}\label{p22}
\Psi_+(\lambda)=\Psi_-(\lambda){\Eu S}_k,\quad
\lambda\in\gamma_k,
\end{equation}
where
\begin{equation}\label{p23}
{\Eu S}_{2k-1}=\pmatrix{1&0\cr s_{2k-1}& 1\cr},\quad
{\Eu S}_{2k}=\pmatrix{1& s_{2k}\cr 0& 1\cr },
\end{equation}
and the parameters $s_k$ do not depend neither on $x$ nor on $\lambda$ and 
satisfy the constraints
\begin{equation}\label{p24}
s_{k+3}=-s_k,\quad
s_1-s_2+s_3+s_1s_2s_3=0.
\end{equation}
Here, $\Psi_+(\lambda)$ and $\Psi_-(\lambda)$ are the respective limits of 
$\Psi(\lambda)$ to the left and to the right of $\gamma$.
\end{description} 
The RH problem is depicted in Figure 1. 

\begin{figure}[hbt]\label{pf1}
\begin{center}
\mbox{\psbox{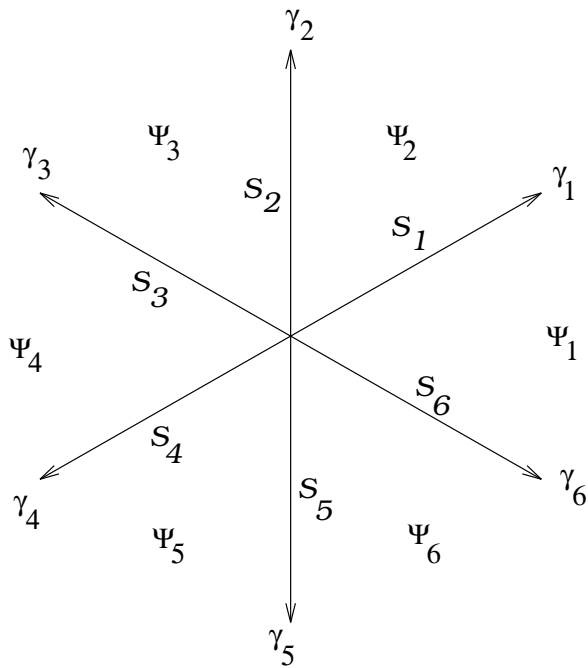}}
\caption{The RH problem graph $\gamma$ for the $\Psi$-function.}
\end{center}
\end{figure}

It is worth noticing that the first of the relations (\ref{p24}) and the 
central symmetry of the RH problem graph imply the symmetry equation 
\begin{equation}\label{p210a}
\Psi(\lambda)=\sigma_2\Psi(-\lambda)\sigma_2,\quad 
\sigma_2=\pmatrix{0&-i\cr i& 0\cr}.
\end{equation}
Also, the special triangular structure of the matrices ${\Eu S}_k$ 
ensures that the product 
$e^{-\theta\sigma_3}{\Eu S}_ke^{\theta\sigma_3}\to I$ as $\lambda\to\infty$,
i.e.\ the well-posedness of the RH problem. 

The piecewise holomorphic function $\Psi(\lambda)$ is in fact a collection of 
the canonical fundamental solutions of the linear $2\times2$ matrix ODE
\begin{equation}\label{p25}
\frac{d\Psi}{d\lambda}=\bigl\{-i\bigl(4\lambda^2+x+2u^2\bigr)\sigma_3-
4u\lambda\sigma_2-2v\sigma_1\bigr\}\Psi,
\end{equation}
$$
\sigma_1=\pmatrix{0&1\cr 1& 0\cr},$$
where $v$ and $u$ are (meromorphic) functions of $x$ such that $v=u_x$ and 
$u(x)$ satisfies the second Painlev\'e equation (\ref{p2}). The jump matrices 
${\Eu S}_k$ of the RH problem have the meaning of the Stokes matrices for 
(\ref{p25}) and, by construction, do not depend on $x$. Therefore, the second 
Painlev\'e equation (\ref{p2}) describes the isomonodromy deformations of 
system (\ref{p25}).

The Painlev\'e function $u(x)$ admits the following representation in terms 
of the solution $\Psi(\lambda)$ of the RH problem,
\begin{equation}\label{nonlinintegral}
u(x)=2\lim_{\lambda\to\infty}
\Bigl(\lambda\Psi_{12}(\lambda)e^{-\theta(\lambda,x)}\Bigr).
\end{equation} 
Moreover, the respective map,
$$
\{s\}\equiv\{s_{k}\colon
s_{k+3}=-s_k,\quad
s_1-s_2+s_3+s_1s_2s_3=0\}\mapsto\{\mbox{solutions of (\ref{p2})}\},$$
is a bijection. Hence we adopt the notation
$$
u(x)\equiv 
u(x|s)$$
for the second Painlev\'e transcendent. 

The following important symmetry properties take place,
\begin{equation}\label{sym0}
u(x|s)=e^{i\frac{2\pi}{3}l}u(xe^{i\frac{2\pi}{3}l}|s^{2l}),
\end{equation}
\begin{equation}\label{sym00}
u(x|s)=\bar{u}(\bar{x}|s^{*}),
\end{equation}
where $s\mapsto s^{d}$, $d\in{\Bbb Z}$ and $s\mapsto s^{*}$ denote the
automorphisms of the set of monodromy data $\{s\}$ defined by the relations 
$$
(s^{d})_{j}:=s_{j+d},$$
and
$$
(s^{*})_{j}:=\bar{s}_{4-j}.$$
Symmetries (\ref{sym0}) and (\ref{sym00}) show that to obtain a complete
asymptotic description of the function $u(x|s)$ as $x\to\infty$, 
$x\in{\Bbb C}$, it is enough to find, in terms of $s$, the asymptotic 
behavior just in one of the sectors 
$\bigl(\frac{\pi}{3}k,\frac{\pi}{3}(k+1)\bigr)$, say in the sector 
$\bigl(\frac{2\pi}{3},\pi\bigr)$. The asymptotics of $u(x|s)$ in this sector 
is then of our principal interest. 

Let us introduce a complex valued function $\varkappa=\varkappa(\varphi)$,
$0<\varphi<\frac{\pi}{3}$, which satisfies the system of transcendent 
equations
\begin{eqnarray}\label{boutroux_equation_KK'}
&&\Re\Bigl[\Bigl(\frac{e^{i\varphi}}{1+\varkappa^2}\Bigr)^{3/2}
\bigl(-2\varkappa^2K'+(1+\varkappa^2)E'\bigr)\Bigr]=0,\nonumber
\\
&&\Im\Bigl[\Bigl(\frac{e^{i\varphi}}{1+\varkappa^2}\Bigr)^{3/2}
\bigl((\varkappa^2-1)K+(1+\varkappa^2)E\bigr)\Bigr]=0,
\end{eqnarray}
where $K$ and $E$ are the standard complete elliptic integrals
$$
K=K(\varkappa)=\int_0^1\frac{dz}{\sqrt{(1-z^2)(1-\varkappa^2z^2)}},\quad
K'=K(\varkappa'),$$
$$
E=E(\varkappa)=\int_0^1\sqrt{\frac{1-\varkappa^2z^2}{1-z^2}}\,dz,\quad
E'=E(\varkappa'),$$
$$
\varkappa'=\sqrt{1-\varkappa^2},\quad
0<\arg\varkappa'<\frac{\pi}{2},\quad
-\frac{\pi}{2}<\arg(1+\varkappa^2)<0.$$
One can show that system (\ref{boutroux_equation_KK'}) has a unique smooth 
solution normalized by the conditions,
$$
\phi\rightarrow\frac{2\pi}{3}\Rightarrow\varkappa\rightarrow1,
\qquad 
\phi\rightarrow\pi\Rightarrow\varkappa\rightarrow0.$$
$$
-\frac{\pi}{2}<\arg\varkappa<0.$$
The main result concerning the asymptotics of $u(x|s)$ for complex $x$ can be 
now formulated as the following theorem.

\smallskip
\begin{theorem}\label{theorem1}
Let
\begin{equation}\label{condition_of_general_position}
s_3\neq0,\quad
1-s_1s_3\neq0.
\end{equation}
Then
\begin{eqnarray}\label{y_sn}
&&u(x|s)=-\frac{i\varkappa}{\sqrt{1+\varkappa^2}}\,
x^{1/2}\sn\Bigl(
\frac{2i}{3}\,\frac{x^{3/2}}{\sqrt{1+\varkappa^2}}-
\frac{2iK}{\pi}\ln(s_3)-\nonumber
\\
&&\hskip3.5cm
+\frac{K'}{\pi}\ln(1-s_1s_3)+
{\cal O}\bigl(x^{-3/2}\bigr)\bigr|\varkappa\Bigr),
\\
&&x\to\infty,\quad
\arg x=\varphi+{\cal O}\bigl(x^{-3/2}\bigr),\quad
\varphi\in\bigl(\frac{2\pi}{3},\pi\bigr),
\end{eqnarray}
and the asymptotic is uniform on the cheese-type domain
$$
\bigl\{x\in{\Bbb C}\colon\quad
|\arg x-\varphi|<C|x|^{-3/2}\bigr\}\cap D_{\varepsilon},$$
where $D_{\varepsilon}$ is the complement to the union of the 
$\varepsilon$-neighborhoods of all the poles of the indicated elliptic 
function.
\end{theorem}

This theorem is due to  V. Novokshenov \cite{nov1} and A. Kapaev 
\cite{kapaev1}. Formulae (\ref{boutroux_equation_KK'}) and (\ref{y_sn}) 
(without the indicated expression for the phase) constitute the classical 
Boutroux ansatz. In this paper we present a new proof of this theorem based 
on the Deift-Zhou nonlinear steepest descent method. This proof is free from 
any prior assumptions, which were crucial for \cite{nov1} and \cite{kapaev1}, 
including the assumption of the solvability of the RH problem under the 
conditions (\ref{condition_of_general_position}). The latter we will obtain 
{\it en route}.

We complete this section by noticing that Theorem~\ref{theorem1} together 
with the symmetry relations (\ref{sym0}) and (\ref{sym00}) yield the 
following asymptotic description for the rest of the sectors.

\begin{corollary}\label{corollary1}
If $s_{2+2l}\neq0$ and $1+s_{2+2l}s_{3+2l}\neq0$ for $l=0,\pm1$, then the 
leading asymptotic term of the Painlev\'e function in the interior of the 
sector $\arg x\in\bigl(-\frac{2\pi}{3}l;\frac{\pi}{3}-\frac{2\pi}{3}l\bigr)$ 
is elliptic.

If $s_{2+2l}\neq0$ and $1+s_{1+2l}s_{2+2l}\neq0$ for $l=0,\pm1$, then the
leading asymptotic term of the Painlev\'e function in the interior of the 
sector $\arg x\in\bigl(-\frac{\pi}{3}-\frac{2\pi}{3}l;-\frac{2\pi}{3}l\bigr)$ 
is elliptic.

The phase shift of the elliptic asymptotics in each of the sectors is given 
by the expression,
$$
\frac{2iK}{\pi}\ln s_{K}\pm\frac{K'}{\pi}\ln s_{K'},$$
where the choice of the monodromy parameters $s_{K}$ and $s_{K'}$ is 
indicated in Figure~2. Together with the monodromy equation (\ref{p24}) this 
constitutes the connection formulae for the Painlev\'e function $u(x|s)$ in 
the complex domain.
\end{corollary}
\begin{figure}[hbt]\label{pict2}
\begin{center}
\mbox{\psbox{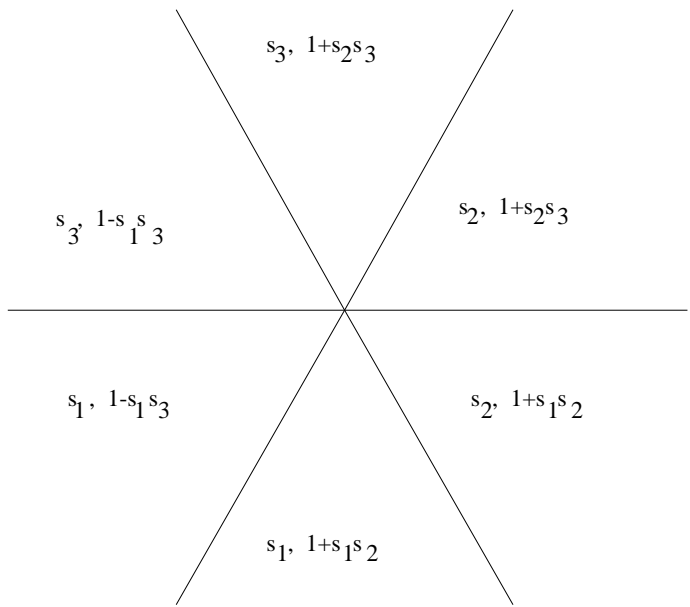}}
\caption{Combinations of the Stokes multipliers whose non-triviality yields
the elliptic asymptotic behavior of Painlev\'e function in
the corresponding sector}
\end{center}
\end{figure}

\begin{@remark}
A complete description of the asymptotic behavior of $u(x|s)$ which includes 
the ``Stokes rays'' $\arg x=\frac{\pi}{3}k$ is given in \cite{kapaev2}. In 
fact, the complete description of the asymptotic behavior of the solutions of 
the second Painlev\'e equation obtained in \cite{kapaev2} covers the general 
case of the latter, i.e.\ the equation
\begin{equation}\label{p2gen}
\frac{d^2u}{dx^2}=2u^3+xu+\alpha,\quad
\alpha\in{\Bbb C}.
\end{equation}
\end{@remark}

\section{The transformation of the RH problem}
\label{RHp_transform}
\setcounter{equation}{0}
Let us scale the independent variable $\lambda$ according to the equation
\begin{eqnarray}\label{p211}
&&\lambda=(e^{-i\varphi}x)^{1/2}z,\quad
(e^{-i\varphi}x)^{3/2}=t,\quad
\varphi=const\in(\frac{2\pi}{3},\pi),
\\
&&\Re t\to+\infty,\quad 
|\Im t|<const.\nonumber
\end{eqnarray}

The stationary points of the phase 
$\theta=it\bigl(\frac{4}{3}z^3+e^{i\varphi}z\bigr)$ are independent of $t$, 
and formally, the problem can be treated in the very same way as the RH 
problem which is dealt with in ref.\ \cite{DZ2, IFK} where the case 
$x\to-\infty$, i.e.\ $\varphi=\pi$, was considered. Indeed, applying 
literally the same arguments as in ref.\ \cite{IFK}, one can easily replace 
rays $\gamma_k$ by the anti-Stokes lines $\gamma_k^{\pm}=\{\Im\theta=0\}$ 
indicated in Figure~3.
\begin{figure}[hbt]\label{pins1}
\begin{center}
\mbox{\psbox{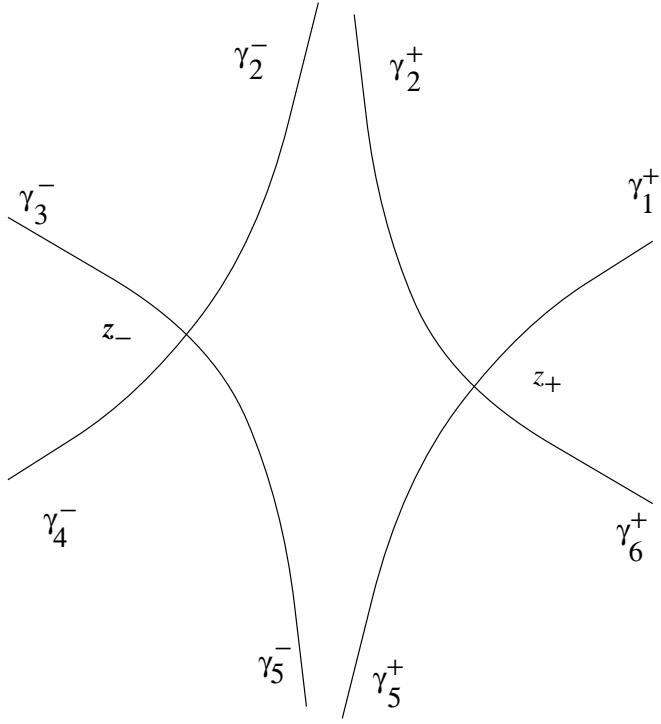}}
\caption{The anti-Stokes lines for the phase
$\theta(z)=it\bigl(\frac{4}{3}z^3+e^{i\varphi}z\bigr)$.}
\end{center}
\end{figure}
Points $z_{\pm}$ are the stationary phase points, 
$z_{\pm}=\mp\frac{i}{2}e^{i\varphi /2}$, and, unlike the case $\varphi=\pi$, 
the exponent $\theta(z)$ has nonzero real parts at $z=z_{\pm}$:
$$
\Re\theta(z_{\pm})\to\pm\frac{t}{3}\cos\frac{3\varphi}{2},\quad
t\to+\infty;$$
since $\varphi\in(\frac{2}{3}\pi,\pi)$, for sufficiently large $t$,
\begin{equation}\label{ins1}
\Re\theta(z_+)<0,\quad
\Re\theta(z_-)>0.
\end{equation}
Unlike the case $\varphi=\pi$, now,
after the normalization of the RH problem (i.e. 
$\Psi \to Y = \Psi e^{\theta(z) \sigma_{3}}$), only {\it half} of the jump matrices,
namely the matrices corresponding to the lines $\gamma_1^+$, $\gamma_2^-$,
$\gamma_4^-$, $\gamma_5^+$, approach the identity as $t\to\infty$ 
{\it uniformly} for $z\in\gamma_k^{\pm}$. The other half of the jump matrices,
i.e.\ the matrices associated with the lines $\gamma_2^+$, $\gamma_3^-$,
$\gamma_5^-$, $\gamma_6^+$, ``explode'' as $t\to\infty$ in the finite
neighborhoods of the points $z_{\pm}$. Therefore, we have no reason to expect 
that the RH problem under consideration can be reduced to the Weber-Hermite 
model RH problems associated to the stationary phase points $z_{\pm}$. Thus, 
the principal idea of the asymptotic analysis in the case $\varphi=\pi$
fails as $\varphi\in(\frac{2\pi}{3},\pi)$.

A regular procedure that allows us to overcome the indicated above obstacles
is based on the following construction. Firstly, we ``split" the stationary
phase points $z_{\pm}$ into four new points
$$
z_1,\ z_2=-z_1,\ z_3,\ z_4=-z_3,$$
where $z_1$, $z_3$ lie in a neighborhood of the stationary phase point $z_+$,
while $z_2$, $z_4$ are associated in the same way with the stationary point
$z_-$. Secondly, we assume that there exists an analytical function $g(z)$,
which satisfies the conditions:

a) $g(z)=i\bigl(\frac{4}{3}z^3+e^{i\varphi}z\bigr) +
{\cal O}\bigl(\frac{1}{z}\bigr)$, as $z\to\infty$;

b) points $\pm z_1$, $\pm z_3$ are the branch points of $g'(z)$ of order 2
and, in their neighborhoods,
$$
g'(z)=c_{1,3}^{\mp}(z\pm z_{1,3})^{1/2}+\dots \,,$$
with $c_{1,3}^{\mp}\neq 0$.

c) $\Re g(\pm z_{1,3})=0.$

d) $g'\neq 0$ on ${\Bbb C}\backslash \{ \pm z_{1,3}\} $.

In fact, conditions a)--d) constitute a certain problem of the theory of 
analytic functions. Simultaneously, they are the restrictions on the points 
$\pm z_{1,3}$ themselves. In Section~\ref{g_function}, we will solve the 
problem by the use of an elliptic integral associated with the elliptic curve 
$w^2=(z^2-z_1^2)(z^2-z_3^2)$. We see then that condition c) is just the 
classical Boutroux equations, which always appear in the asymptotic analysis 
of the Painlev\'e equations. It also should be emphasizing that a flexibility 
we gain introducing four parameters, i.e.\ points $\pm z_{1,3}$, plays the 
crucial role for solvability of the problem a)--d).

Having assumed the existence of the function $g(z)$, our main idea is to 
replace the original phase 
$\theta=it\bigl(\frac{4}{3}z^3+e^{i\varphi}z\bigr)$ by the function $g(z)$.
Because of the asymptotic equation~a), this replacement preserves the
normalization condition (\ref{p21}). Moreover, due to condition~c), we
eliminate the main trouble, i.e.\ inequalities (\ref{ins1}). Nevertheless,
there is a price: now, we have to deform the initial RH problem to the RH
problem formulated for the anti-Stokes lines $\hat\gamma^{(k)}$ corresponding 
to the function $g(z)$:
\begin{equation}\label{ins2}
\Im g(z)=\Im g(z_k).
\end{equation}
The anti-Stokes graph defined in (\ref{ins2}) is more complicated than the
$\theta$-graph depicted in Figure~3. Properties~a), b) and~d) lead to the
$g$-anti-Stokes lines $\hat\gamma_s^{(k)}$ as shown in Figure~4. Our main task
for this section, is the transformation of the initial RH graph presented in
Figure~1 into the one given in Figure~4.

\begin{figure}[hbt]\label{pf2}
\begin{center}
\mbox{\psbox{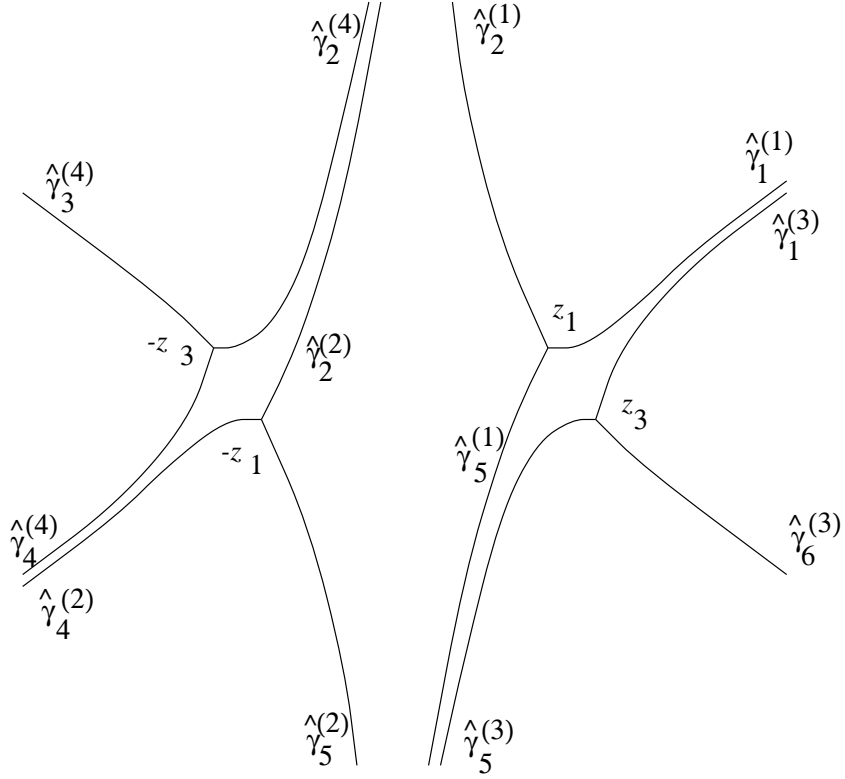}}
\caption{The anti-Stokes lines $\hat\gamma$ for the equation for $x\to\infty$,
$\arg x\in(\frac{2\pi}{3},\pi)$.}
\end{center}
\end{figure}
The notation $\hat\gamma_s^{(k)}$ for the anti-Stokes line used here refers to
its starting point $z_k$ and to the anti-Stokes ray $\gamma_s$ as being 
asymptotic for the line considered.

As it is easy to see, the transformation problem is not trivial and can not be
reduced to the simple bending of the graph branches due to the very different
topological properties of the graphs presented in Figure~1 and Figure~4. In
particular, the RH problem graph $\gamma$ for the $\Psi$ function is 
connected, while the anti-Stokes graph $\hat\gamma$ is disconnected. The most 
we can do with the original problem by means of trivial bending of the 
branches is shown in Figure~5.
\begin{figure}[hbt]\label{pf3}
\begin{center}
\mbox{\psbox{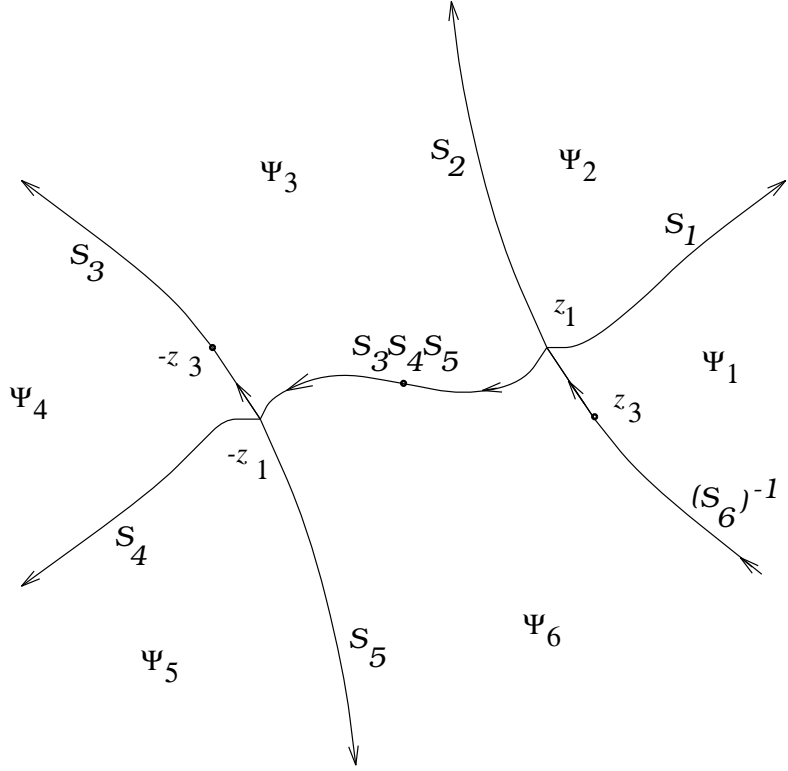}}
\caption{The continuously deformed RH problem graph.}
\end{center}
\end{figure}

Note that after such deformation, a new boundary curve between the third and 
the sixth domains arises where the jump is described by the ratio of the 
sixth and the third canonical solutions:
\begin{equation}\label{p212}
\Psi_6(z)=\Psi_3(z){\Eu S}_3{\Eu S}_4{\Eu S}_5,
\end{equation}
where
$$
{\Eu S}_3{\Eu S}_4{\Eu S}_5=
\pmatrix{1+s_1s_2&-s_1\cr -s_1 & 1-s_1s_3\cr},$$

Let us consider the function $\delta(z)$ depending on $z_k$, $k=1,3$, 
and some parameters $\nu_l$, $l=1,2,3$,
\begin{equation}\label{p213}
\delta(z)=
\Bigl(\frac{z+z_1}{z+z_3}\Bigr)^{\nu_3}
\Bigl(\frac{z-z_1}{z+z_1}\Bigr)^{\nu_2}
\Bigl(\frac{z-z_3}{z-z_1}\Bigr)^{\nu_1},
\end{equation}
which is defined on the complex plane cut along the broken line
$[z_3,z_1]\cup[z_1,-z_1]\cup[-z_1,-z_3]$.
The function is fixed by condition
\begin{equation}\label{p214}
\delta(z)\longto_{z\to\infty}1,
\end{equation}
and has the jumps on the cuts
\begin{eqnarray}\label{p215}
&&\delta_-(z)=\delta_+(z)e^{2\pi i\nu_1},\quad
z\in(z_3,z_1),\nonumber
\\
&&\delta_-(z)=\delta_+(z)e^{2\pi i\nu_2},\quad
z\in(z_1,-z_1),
\\
&&\delta_-(z)=\delta_+(z)e^{2\pi i\nu_3},\quad
z\in(-z_1,-z_3),\nonumber
\end{eqnarray}
as it is shown in Figure~6.

\begin{figure}[hbt]\label{pf4}
\begin{center}
\mbox{\psbox{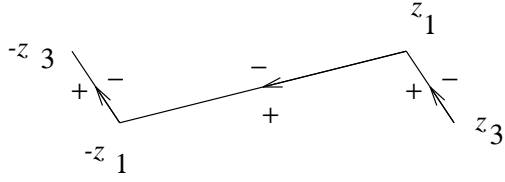}}
\caption{The graph of the scalar RH problem solved by the 
function $\delta(z)$.}
\end{center}
\end{figure}

The next step is the most important for the transformation of the original
RH problem. Let us introduce the new function
\begin{equation}\label{p216}
\hat\Psi=\Psi\delta^{-\sigma_3}.
\end{equation}
This new function does not differ from the original one at infinity, but its
jumps are described by the new matrices:
\begin{equation}\label{p217}
\hat\Psi_+=\hat\Psi_-\delta_-^{\sigma_3}{\Eu S}\delta_+^{-\sigma_3}=
\hat\Psi_-\hat{\Eu S}.
\end{equation}
The new RH graph is shown in Figure~7, and the new jump matrices 
$\hat{\Eu S}$ are given by the equations,

\begin{figure}[hbt]\label{pf5}
\begin{center}
\mbox{\psbox{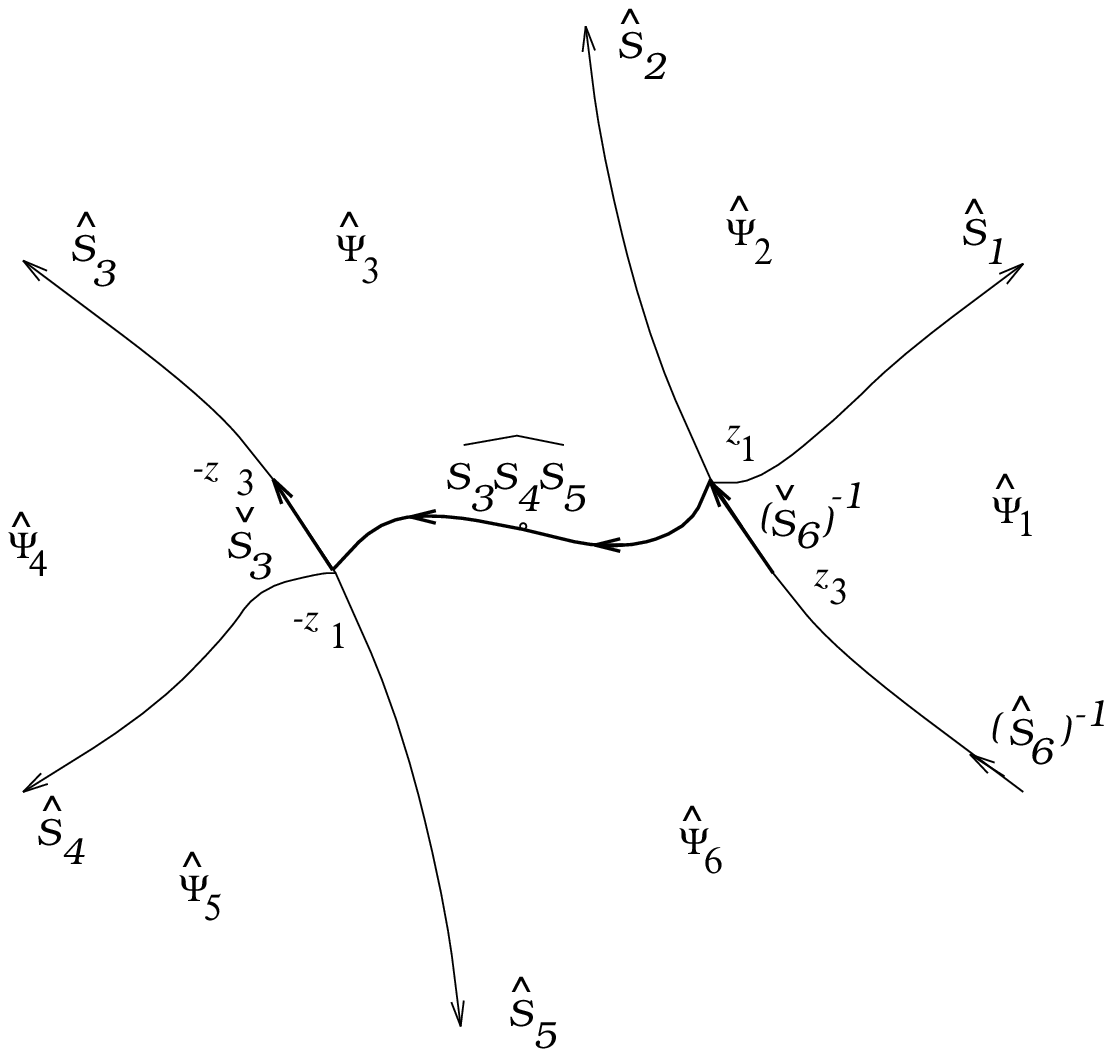}}
\caption{The RH problem for $\hat\Psi$.}
\end{center}
\end{figure}

\begin{eqnarray}\label{p218}
&&\hat{\Eu S}_{2k-1}=\pmatrix{1&0\cr s_{2k-1}\delta^{-2}&1\cr },\quad
\hat{\Eu S}_{2k}=\pmatrix{1& s_{2k}\delta^2\cr 0&1\cr },
\\
&&\check{\Eu S}_3=\pmatrix{e^{2\pi i\nu_3}&0\cr
s_3\delta_-^{-2}e^{2\pi i\nu_3} &e^{-2\pi i\nu_3}\cr},\quad
\bigl(\check{\Eu S}_6\bigr)^{-1}=
\pmatrix{e^{2\pi i\nu_1}&s_3\delta_-^2e^{-2\pi i\nu_1}\cr
0 &e^{-2\pi i\nu_1}\cr},\nonumber
\\
&&\HAT{{\Eu S}_3{\Eu S}_4{\Eu S}_5}=
\pmatrix{(1+s_1s_2)e^{2\pi i\nu_2}&-s_1\delta_-^2e^{-2\pi i\nu_2}\cr
-s_1\delta_-^{-2}e^{2\pi i\nu_2}&
(1-s_1s_3)e^{-2\pi i\nu_2}\cr}\,.\nonumber
\end{eqnarray}

Now, we are going to explain what this  change is made for. 
The point is that some
of the new jump matrices can be factorized in such a way, that the simple
transformation of the RH graph allows us to approach the anti-Stokes lines 
indicated in Figure~4 (cf. \cite{IFK}).
Indeed, let us assume that the generic conditions hold 
true,
\begin{equation}\label{generic}
s_3\neq0,\quad
1-s_1s_3\neq0.
\end{equation}
If we then put
\begin{equation}\label{p219}
e^{-2\pi i\nu_3}=s_3,
\end{equation}
then it is easy to check the equation holds:
\begin{equation}\label{p220}
\check{\Eu S}_3=\hat{\Eu S}_2^U\hat{\Eu S}_A\hat{\Eu S}_4^U\,,
\end{equation}
where
\begin{equation}\label{p221}
\hat{\Eu S}_2^U=\pmatrix{1&\delta_-^2s_3^{-1}\cr 0&1\cr}\,,\quad
\hat{\Eu S}_A=\pmatrix{ &-\delta_-^2\cr \delta_-^{-2}& \cr }\,,\quad
\hat{\Eu S}_4^U=\pmatrix{1&\delta_+^2s_3^{-1}\cr 0&1\cr}\,.
\end{equation}
Similarly, if we put
\begin{equation}\label{p222}
e^{2\pi i\nu_1}=s_3\,,\quad\hbox{or}\quad
\nu_1=-\nu_3\,,
\end{equation}
then the factorization arises
\begin{equation}\label{p223}
\check{\Eu S}_6^{-1}=\hat{\Eu S}_1^L\hat{\Eu S}_B\hat{\Eu S}_5^L\,,
\end{equation}
with the matrices
\begin{equation}\label{p224}
\hat{\Eu S}_1^L=\pmatrix{1& 0\cr (\delta_-^2s_3)^{-1}&1\cr}\,,\quad
\hat{\Eu S}_B=\pmatrix{ &\delta_-^2\cr -\delta_-^{-2}& \cr}\,,\quad
\hat{\Eu S}_5^L=\pmatrix{1& 0\cr (\delta_+^2s_3)^{-1}&1\cr}\,.
\end{equation}
At the same time, if we put
\begin{equation}\label{p225}
e^{2\pi i\nu_2}=1-s_1s_3\,,
\end{equation}
then
\begin{equation}\label{p226}
\HAT{{\Eu S}_3{\Eu S}_4{\Eu S}_5}=\hat{\Eu S}^U\hat{\Eu S}^L
\end{equation}
with
\begin{equation}\label{p227}
\hat{\Eu S}^U=\pmatrix{1& -\frac{s_1}{1-s_1s_3}\delta_-^2\cr 0& 1\cr}\,,\quad
\hat{\Eu S}^L=\pmatrix{1& 0\cr
-\frac{s_1}{1-s_1s_3}\delta_+^{-2}& 1\cr}\,.
\end{equation}

The ``intermediate" RH graph is shown in Figure~8, and the final
graph is presented in Figure~9.

\begin{figure}[hbt]\label{pf6}
\begin{center}
\mbox{\psbox{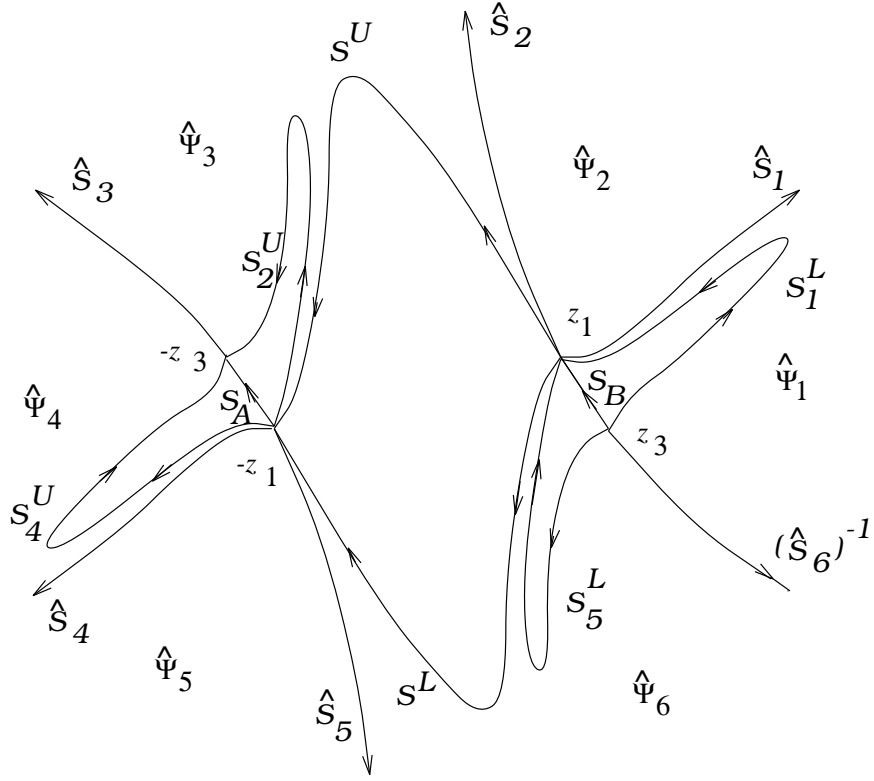}}
\caption{Deformation of the RH problem graph for $\hat\Psi$.}
\end{center}
\end{figure}

\begin{figure}[hbt]\label{pf7}
\begin{center}
\mbox{\psbox{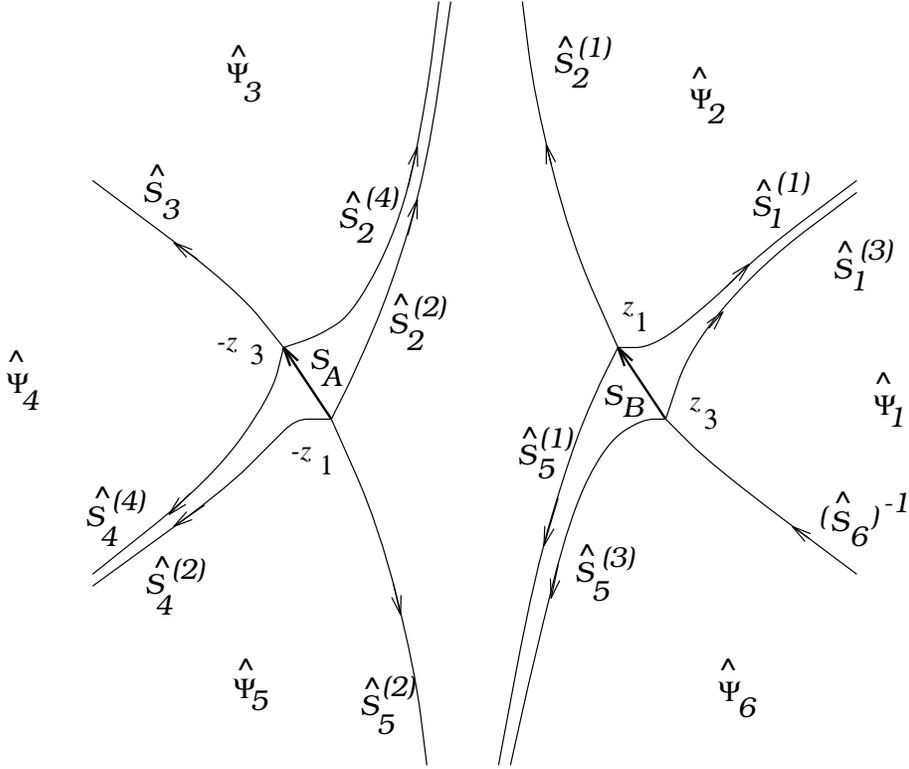}}
\caption{The final RH problem graph for $\hat\Psi$.}
\end{center}
\end{figure}

Here, the jump matrices are
\begin{equation}\label{p230}
\hat{\Eu S}_5^{(3)}=
\pmatrix{{\scr 1}& {\scr 0}\cr \frac{1}{\delta^2s_3}&{\scr 1}\cr},\quad
\hat{\Eu S}_6^{-1}=
\pmatrix{{\scr 1}&{\scr \delta^2s_3}\cr {\scr 0}&{\scr 1}\cr},\quad
\hat{\Eu S}_1^{(3)}=
\pmatrix{{\scr 1}& {\scr 0}\cr \frac{1}{\delta^2s_3}&{\scr 1}\cr},
\end{equation}
\smallskip
\begin{equation}\label{p231}
\hat{\Eu S}_1^{(1)}=
\pmatrix{{\scr 1}& {\scr 0}\cr -\frac{1-s_1s_3}{\delta^2s_3}&{\scr 1}\cr},\
\hat{\Eu S}_2^{(1)}=
\pmatrix{{\scr 1}&\frac{\delta^2s_3}{1-s_1s_3}\cr {\scr 0}&{\scr 1}\cr},\
\hat{\Eu S}_5^{(1)}=
\pmatrix{{\scr 1}& {\scr 0}\cr -\frac{1}{\delta^2s_3(1-s_1s_3)}&{\scr 1}\cr},
\end{equation}
\smallskip
\begin{equation}\label{p233}
\hat{\Eu S}_4^{(2)}=
\pmatrix{{\scr 1}&{\scr \delta^2}\frac{1-s_1s_3}{s_3}\cr 
{\scr 0}&{\scr 1}\cr},\
\hat{\Eu S}_5^{(2)}=
\pmatrix{{\scr 1}&{\scr 0}\cr -\frac{s_3}{\delta^2(1-s_1s_3)}&{\scr 1}\cr},\
\hat{\Eu S}_2^{(2)}=
\pmatrix{{\scr 1}&\frac{\delta^2}{s_3(1-s_1s_3)}\cr {\scr 0}&{\scr 1}\cr},
\end{equation}
\smallskip
\begin{equation}\label{p234}
\hat{\Eu S}_2^{(4)}=
\pmatrix{{\scr 1}&-\frac{\delta^2}{s_3}\cr {\scr 0}&{\scr 1}\cr},\quad
\hat{\Eu S}_3=
\pmatrix{{\scr 1}&{\scr 0}\cr \frac{s_3}{\delta^2}&{\scr 1}\cr},\quad
\hat{\Eu S}_4^{(4)}=
\pmatrix{{\scr 1}&-\frac{\delta^2}{s_3}\cr {\scr 0}&{\scr 1}\cr},
\end{equation}
\smallskip
\begin{equation}\label{p235}
\hat{\Eu S}_A=\pmatrix{ &-{\scr \delta_-^2}\cr \frac{1}{\delta_-^2}& \cr},\quad
\hat{\Eu S}_B=\pmatrix{ &{\scr \delta_-^2}\cr -\frac{1}{\delta_-^2}& \cr}.
\end{equation}
Also, taking into account the equations (\ref{p219}), (\ref{p222}) and
(\ref{p225}), we rewrite the jump conditions (\ref{p215}) for the function 
$\delta(z)$ (\ref{p213}) as follows,
\begin{eqnarray}\label{p236}
&\delta_-(z)=\delta_+(z)s_3,\quad
&z\in (z_3,z_1),\nonumber
\\
&\delta_-(z)=\delta_+(z)(1-s_1s_3),\quad
&z\in (z_1,-z_1),
\\
&\delta_-(z)=\delta_+(z)\cdot\frac{1}{s_3},\quad
&z\in (-z_1,-z_3).\nonumber
\end{eqnarray}

In this form, the RH problem graph for the  $\hat\Psi$-function
coincides with the anti-Stokes graph $\hat\gamma$, see Figure~4, except for 
the additional segments $[z_3,z_1]$ and~$[-z_1,-z_3]$. 

Now, we do not need the function $\delta$ anymore and can return to the  
original $\Psi$-function by the use of (\ref{p216}), 
\begin{equation}\label{p216a}
\Psi=\hat\Psi\delta^{\sigma_3}.
\end{equation}
The calculations above show that the solution of the second Painlev\'e
equation can be obtained from the $\Psi$-function with the jumps on the graph 
shown on Figure~10. The new RH problem, which replaces
the problem  (\ref{p21})--(\ref{p24}), can be
formulated as follows.
\begin{figure}[hbt]\label{pf7a}
\begin{center}
\mbox{\psbox{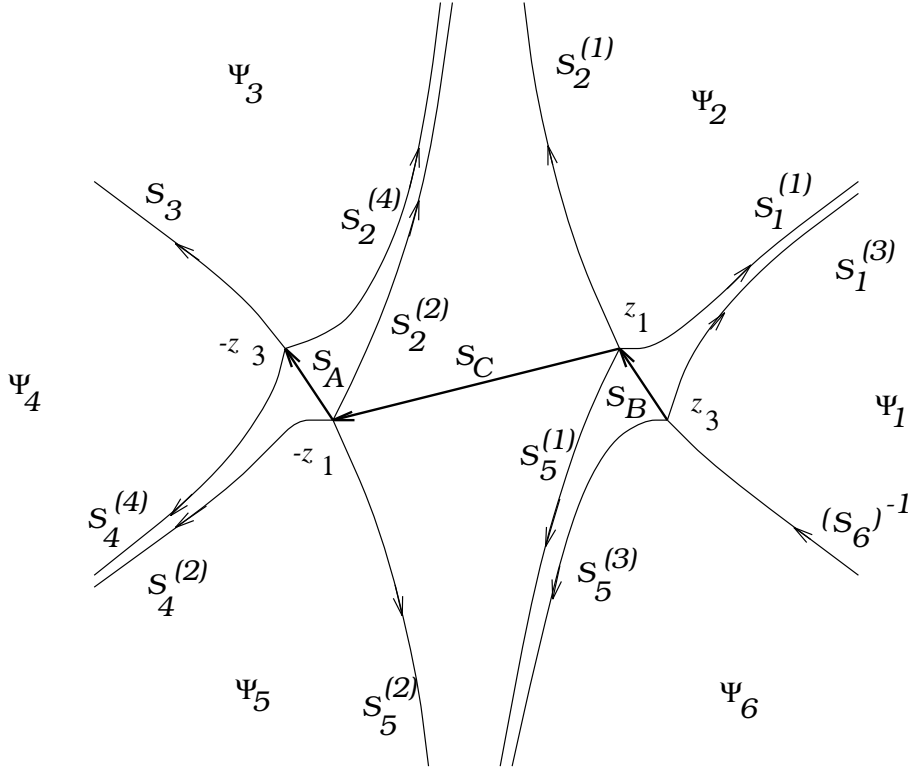}}
\caption{The ultimate RH problem graph for $\Psi$.}
\end{center}
\end{figure}

\begin{description}
\item{1)} Find a piecewise holomorphic matrix-valued $2\times2$ function
$\Psi$ of the complex variable $z$ with an irregular singular point at
infinity where it has the exponential asymptotic behavior of the type
(\ref{p21})
\begin{equation}\label{p21u}
\Psi(z)=\bigl(I+{\cal O}(z^{-1})\bigr)
e^{-t\theta(z)\sigma_3},\quad
\theta(z)=i\bigl(\frac{4}{3}z^3+e^{i\varphi}z\bigr),\quad
z\to\infty;
\end{equation}
\item{2)} on the piecewise smooth oriented graph $\hat\gamma$, see Figure~10, 
the jump condition holds true,
\begin{equation}\label{p22u}
\Psi_+(z)=\Psi_-(z){\Eu S},\quad
\lambda\in\hat\gamma.
\end{equation}
Here, $\Psi_+(z)$ and $\Psi_-(z)$ are the limits of the function $\Psi(z)$ on 
the contour $\hat\gamma$ to the left and to the right, respectively, and the 
piecewise constant matrix ${\Eu S}$ is described  by the following equations,
\begin{equation}\label{p230u}
{\Eu S}_5^{(3)}=
\pmatrix{{\scr 1}& {\scr 0}\cr \frac{1}{s_3}&{\scr 1}\cr},\quad
{\Eu S}_6^{-1}=
\pmatrix{{\scr 1}&{\scr s_3}\cr {\scr 0}&{\scr 1}\cr},\quad
{\Eu S}_1^{(3)}=
\pmatrix{{\scr 1}& {\scr 0}\cr \frac{1}{s_3}&{\scr 1}\cr},
\end{equation}
\smallskip
\begin{equation}\label{p231u}
{\Eu S}_1^{(1)}=
\pmatrix{{\scr 1}& {\scr 0}\cr -\frac{1-s_1s_3}{s_3}&{\scr 1}\cr}\!,\
{\Eu S}_2^{(1)}=
\pmatrix{{\scr 1}&\frac{s_3}{1-s_1s_3}\cr {\scr 0}&{\scr 1}\cr}\!,\
{\Eu S}_5^{(1)}=
\pmatrix{{\scr 1}& {\scr 0}\cr -\frac{1}{s_3(1-s_1s_3)}&{\scr 1}\cr}\!,
\end{equation}
\smallskip
\begin{equation}\label{p233u}
{\Eu S}_4^{(2)}=
\pmatrix{{\scr 1}&\frac{1-s_1s_3}{s_3}\cr 
{\scr 0}&{\scr 1}\cr},\
{\Eu S}_5^{(2)}=
\pmatrix{{\scr 1}&{\scr 0}\cr -\frac{s_3}{1-s_1s_3}&{\scr 1}\cr},\
{\Eu S}_2^{(2)}=
\pmatrix{{\scr 1}&\frac{1}{s_3(1-s_1s_3)}\cr {\scr 0}&{\scr 1}\cr},
\end{equation}
\smallskip
\begin{equation}\label{p234u}
{\Eu S}_2^{(4)}=
\pmatrix{{\scr 1}&-\frac{1}{s_3}\cr {\scr 0}&{\scr 1}\cr},\quad
{\Eu S}_3=
\pmatrix{{\scr 1}&{\scr 0}\cr {\scr s_3}&{\scr 1}\cr},\quad
{\Eu S}_4^{(4)}=
\pmatrix{{\scr 1}&-\frac{1}{s_3}\cr {\scr 0}&{\scr 1}\cr},
\end{equation}
\smallskip
\begin{equation}\label{p235u}
{\Eu S}_A=\pmatrix{ &-\frac{1}{s_3}\cr {\scr s_3}& \cr},\quad
{\Eu S}_B=\pmatrix{ &{\scr s_3}\cr -\frac{1}{s_3}& \cr},\quad
{\Eu S}_C=\pmatrix{\frac{1}{1-s_1s_3}&\cr &{\scr 1-s_1s_3}\cr}.
\end{equation}
The parameters $s_k$ satisfy the constraint (\ref{p24}),
\begin{equation}\label{p24u}
s_1-s_2+s_3+s_1s_2s_3=0,
\end{equation}
and do not depend neither on $t$ nor on $z$. 
\end{description}

\begin{@remark}
Possibly, given the jump contour $\hat\gamma$, the resulting jump conditions 
(\ref{p230u})--(\ref{p235u}) can be reconstructed directly from the jump 
conditions (\ref{p23}) for the contour $\gamma$, see Figure~1, using, instead 
of the ``ghost'' function $\delta(z)$, the system of algebraic relations for 
the matrices ${\Eu S}_k^{(l)}$ and ${\Eu S}_k$.
\end{@remark}

\section{Construction of the function $g(z)$}
\label{g_function}
\setcounter{equation}{0}

In this section, we construct the function $g(z)$ described in Section~3.
Consider the function 
\begin{equation}\label{p31}
w(z)=\sqrt{(z^2-z_1^2)(z^2-z_3^2)},\quad
z_1^2+z_3^2=-\frac{1}{2}e^{i\varphi},
\end{equation}
defined on the Riemann surface $\Gamma$, see Figure~11, pasted of two complex 
planes cut along the intervals $[z_3,z_1]$ and $[-z_1,-z_3]$, while
the branch of the root on the upper sheet is chosen in such a way that
$$
w=z^2-\frac{1}{2}\bigl(z_1^2+z_3^2\bigr)+
{\cal O}\bigl(z^{-2}\bigr)\quad\hbox{as}\quad
z\to\infty^+.$$
Here $\infty^+$ denotes the infinite point of the upper sheet. We choose the 
basis $\{a,b\}$ of the group $H_1(\Gamma)$ as it is indicated in Figure~11.
\begin{figure}[hbt]\label{pf8}
\begin{center}
\mbox{\psbox{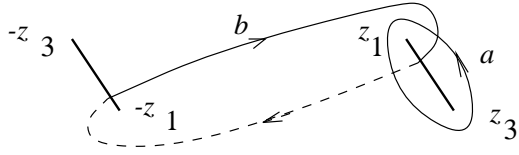}}
\caption{The Riemann surface of the function $w(z)$. $a$ and $b$
mean the basic cycles of the surface.}
\end{center}
\end{figure}
Introduce the Abelian (elliptic, in fact) integral of the second kind
\begin{equation}\label{p32}
\hat g(z)=4i\int_{0_+}^{z}w(z)\,dz.
\end{equation}
We shall consider $\hat g(z)$ as a single valued function defined on the 
complex plane ${\Bbb C}^*$ cut along the sum of the intervals 
$[z_3,z_1]\cup[z_1,-z_1]\cup[-z_1,-z_3]$, moreover the starting point $z=0_+$ 
of the integration path lies on the left side of the oriented cut 
$[z_1,-z_1]$. Introduce the periods of (\ref{p32}),
\begin{equation}\label{p32a}
A=4i{\cal J}_a=4i\oint_{a}w(z)\,dz\,,\quad
B=4i{\cal J}_b=4i\oint_{b}w(z)\,dz\,.
\end{equation}
Taking into account the symmetry $w(-z)=w(z)$, we
find that
$$
\hat g(z)+\hat g(-z)=
4i\Bigl(\int_{0_+}^{z}+
\int_{0_+}^{z_1}+\oint_a+\int_{z_1}^{0_-}+\int_{0_-}^{-z}\Bigr)
w(z)\,dz=A.$$
Thus the function
\begin{equation}\label{p34}
g(z)=\hat g(z)-\frac{1}{2}A,
\end{equation}
considered on ${\Bbb C}^*$, is odd, 
$$
g(-z)=-g(z).$$
Using definition (\ref{p34}), we see that properties b) and d) of $g(z)$ are 
satisfied.

To find the asymptotics of $g(z)$ as $z\to\infty$, introduce the regularized 
integral
$$
4i\int_{0_+}^z\bigl(w(z)-z^2+\frac{1}{2}(z_1^2+z_3^2)\bigr)\,dz=
\hat g(z)-i\bigl(\frac{4}{3}z^3+e^{i\varphi}z\bigr).$$
On the other hand, again using the symmetry $w(z)=w(-z)$, it is equal to
$$
2i\Bigl(\int_{-z}^{0_-}+\int_{0_+}^z\Bigr)
\bigl(w(z)-z^2+\frac{1}{2}(z_1^2+z_3^2)\bigr)\,dz=$$
$$
=2i\oint_a\bigl(w(z)-z^2+\frac{1}{2}(z_1^2+z_3^2)\bigr)\,dz+{\cal O}(z^{-1})=
\frac{1}{2}A+{\cal O}(z^{-1}).$$
Thus,
\begin{equation}\label{p35}
g(z)=i\bigl(\frac{4}{3}z^3+e^{i\varphi}z\bigr)+
{\cal O}\bigl(z^{-1}\bigr)\quad\hbox{as}\quad
z\to\infty,
\end{equation}
and we have established the property a) of the function $g(z)$.
The property c) of the function $g(z)$, i.e.\ $\Re g(\pm z_{1,3})=0$, is
equivalent to the system of equations $\Re A=\Re B=0$ and thus takes the 
integral form,
\begin{equation}\label{p37}
\Im\oint_{a,b}w(z)\,dz=0.
\end{equation}
Equations (\ref{p37}), known as the {\it Boutroux equations}, 
are the substantial
ingredient of our proof below of the solvability of the RH problem. 

Denote $D=z_1^2z_3^2$ the module corresponding 
to our elliptic curve. We observe that (\ref{p37}) determines
$D$ as  the differentiable function 
$D(\varphi)$ of the angle parameter $\varphi \in  (\frac{2\pi}{3},\pi)$. 
Indeed, the l.h.s.\ of (\ref{p37}) yields the complete set of 
the independent first integrals of the first order nonlinear ODE 
\begin{equation}\label{dD}
\frac{dD}{d\varphi}=-\frac{1}{4\Im(\omega_a\bar\omega_b)}
\bigl(\bar\omega_a(e^{i\varphi}\Omega_a-e^{-i\varphi}\bar\Omega_a)-
\omega_b(e^{i\varphi}\Omega_b-e^{-i\varphi}\bar\Omega_b)\bigr),
\end{equation}
where
$$
\omega_{a,b}=\oint_{a,b}\frac{dz}{w(z)},\quad
\Omega_{a,b}=\oint_{a,b}\frac{z^2\,dz}{w(z)}.$$
Whenever the elliptic curve $w(z)$ does not degenerate, the r.h.s.\ of
(\ref{dD}) is a smooth function of $\varphi$, and, given the initial point,
equation (\ref{dD}) is uniquely solvable. To find the initial point, observe
that equations (\ref{p37}) admit degeneration $z_1=z_3$ for $\varphi=\pi$
and $z_1=0$ for $\varphi=\frac{2\pi}{3}$. At these boundary points, the 
r.h.s.\ of (\ref{dD}) remains continuous which ensures its solvability. The 
classical uniqueness theorem for ODEs breaks down at these boundary points, 
however the only one of the solutions of the differential equation
(\ref{dD}) passing through the initial boundary point satisfies the integral 
equations (\ref{p37}). 

Define the Stokes line for the linear equation (\ref{p25}) emanating from the 
point $z_k$, $k=1,2,3,4$, as the set of the points $z$ satisfying the equation
\begin{equation}\label{p37a}
\Im\int_{z_k}^{z}w(z)\,dz=0.
\end{equation}
Define the Stokes graph as the union of all the Stokes lines. The conditions 
(\ref{p37}) mean that the Stokes graph is connected. On the other hand,
conditions (\ref{p37}) allow us to choose the arcs of the Stokes lines as the 
segments $[z_3,z_1]$, $[z_1,-z_1]$ and $[-z_1,-z_3]$. 

For the following, we need to know the jump conditions of the function 
$g(z)$. These are described by the equations
\begin{eqnarray}\label{g_jumps}
&&z\in(z_3,z_1)\colon\quad
g_++g_-=\frac{1}{2}B;\nonumber
\\
&&z\in(z_1,-z_1)\colon\quad
g_+-g_-=-A;\nonumber 
\\
&&z\in(-z_1,-z_3)\colon\quad
g_++g_-=-\frac{1}{2}B.
\end{eqnarray}

\section{The model Baker-Akhiezer RH problem.}
\label{BA_function}
\setcounter{equation}{0}

We remind that the reason why we  replaced the original RH problem by the one 
presented in Figure 10 is that 
\begin{equation}\label{ooo}
e^{-tg(z)\sigma_{3}}Se^{tg(z)\sigma_{3}} \to I,
\end{equation}
as $|t|\to\infty$ for all $z$ belonging to the infinite branches of the jump 
contour $\hat{\gamma}$ and $z\neq\pm z_{k}$. Indeed, equation (\ref{ooo}) 
means that when we normalize the RH problem, i.e.\ when we pass from the 
function $\Psi$ to the function
$$
Y(z)=\Psi(z)e^{tg(z)\sigma_{3}},$$
so that $Y(z)\to I$ as $z\to\infty$, the $Y$-jump matrices will exponentially 
fast approach the identity matrix as $|t|\to\infty$ everywhere on the infinite
branches of $\hat{\gamma}$. Therefore, one can expect that the main 
contribution into the asymptotics of the solution of the RH problem will be 
made by the truncated RH problem  posed on the sum of the segments 
(see Figure~6) $(z_3,z_1)\cup(z_1,-z_1)\cup(-z_1,-z_3)$. In other words, we 
arrive to the model problem consisting of finding the piecewise holomorphic 
function $\Psi^{(BA)}(z)$ with the following properties:

\smallskip
\noindent
1) near infinity
\begin{equation}\label{p411}
\Psi^{(BA)}(z)=\bigl(I+{\cal O}(z^{-1})\bigr)e^{-t\theta\sigma_3},\quad 
z\to\infty,
\end{equation}
where $\theta=i\bigl(\frac{4}{3}z^3+e^{i\varphi}z\bigr)$;

\noindent
2) on the sum of the oriented contours
$(z_3,z_1)\cup(z_1,-z_1)\cup(-z_1,-z_3)$, the jump conditions hold
\begin{eqnarray}\label{p412}
\hskip-15pt
z\in(-z_1,-z_3)\colon&&\hskip-10pt
\Psi_+^{(BA)}(z)=\Psi_-^{(BA)}(z){\Eu S}_A,\quad
{\Eu S}_A=\pmatrix{&-\frac{1}{s_3}\cr s_3&\cr},
\\\label{p413}
\hskip-15pt
z\in (z_3,z_1)\colon&&\hskip-10pt
\Psi_+^{(BA)}(z)=\Psi_-^{(BA)}(z){\Eu S}_B,\quad
{\Eu S}_B=\pmatrix{&s_3\cr-\frac{1}{s_3}&\cr},
\\\label{p414}
\hskip-15pt
z\in (z_1,-z_1)\colon&&\hskip-10pt
\Psi_+^{(BA)}(z)=\Psi_-^{(BA)}(z){\Eu S}_C,\
{\Eu S}_C=\pmatrix{\frac{1}{1-s_1s_3}&\cr&1-s_1s_3\cr}\!;
\end{eqnarray}

\noindent
3) the function $\Psi^{(BA)}(z)$ has no more than $L_{2}$-integrable 
singularities at the points $\pm z_1$, $\pm z_3$.

\smallskip
We use the superscript $(BA)$ since we are going to construct 
$\Psi^{(BA)}(z)$ in terms of the Baker-Akhiezer functions of the Riemann 
surface $\Gamma$. On this Riemann surface, there exists the only one, up to 
an arbitrary constant multiplier, holomorphic differential $\frac{dz}{w(z)}$ 
with the periods $\omega_{a,b}$. Introduce the canonical Abelian (elliptic) 
integral of the first kind $U(z)$,
\begin{equation}\label{U}
U(z)=\frac{1}{\omega_a}\int_{\infty^+}^z\frac{dz}{w(z)},
\end{equation}
and the Abelian integral $h(z)$ which differs from $g(z)$ (\ref{p34}) in some 
integral of the first kind,
\begin{equation}\label{p415}
h(z)=g(z)+cU(z),
\end{equation}
where the constant $c$ will be determined later. Since 
$U(z)={\cal O}(z^{-1})$ as $z\to\infty$, the integral $h(z)$ has the same
canonical asymptotics as the integral $g(z)$,
\begin{equation}\label{p416}
h(z)=i\bigl(\frac{4}{3}z^3+e^{i\varphi}z\bigr)+{\cal O}(z^{-1}).
\end{equation}
Denote the corresponding complete integrals by the letters $A'$ and $B'$:
\begin{eqnarray}\label{p417}
&&A'=4i\oint_a w(z)\,dz+\frac{c}{\omega_a}\oint_a\frac{dz}{w(z)}
\equiv A+c,\nonumber
\\
&&B'=4i\oint_b w(z)\,dz+\frac{c}{\omega_a}\oint_b\frac{dz}{w(z)}
\equiv B+c\tau,\quad
\tau=\frac{\omega_b}{\omega_a}.
\end{eqnarray}

Let us introduce the ``intermediate" $\Pso^o$-function,
\begin{equation}\label{p418}
\Pso^o(z)=\Psi^{(BA)}(z)e^{t_0h(z)\sigma_3},
\end{equation}
where $t_0$ is the second reserved parameter to be defined later. The 
asymptotics of this function at infinity,
\begin{equation}\label{p419}
\Pso^o(z)=\bigl(I+{\cal O}(z^{-1})\bigr)e^{-(t-t_0)\theta\sigma_3},\quad 
z\to\infty,
\end{equation}
differs from the canonical in the difference $t-t_0$ which replaces the 
original $t$. Its jump properties are similar to those of the function 
$\Psi^{(BA)}$ except for they are described using the new parameters $t_0$, 
$A'$ and $B'$:
\begin{eqnarray}\label{p420}
z\in (-z_1,-z_3)\colon&&
\Pso^o{\!}_+(z)=\Pso^o{\!}_-(z)\hat G_A(z),
\\\label{p421}
z\in (z_3,z_1)\colon&&
\Pso^o{\!}_+(z)=\Pso^o{\!}_-(z)\hat G_B(z),
\\\label{p422}
z\in (z_1,-z_1)\colon&&
\Pso^o{\!}_+(z)=\Pso^o{\!}_-(z)\hat G_C(z),
\end{eqnarray}
where
$$
\hat G_A=\pmatrix{&-s_3^{-1}e^{t_0 B'/2}\cr s_3e^{-t_0 B'/2}&\cr},\quad
\hat G_B=\pmatrix{&s_3e^{-t_0 B'/2}\cr-s_3^{-1}e^{t_0B'/2}&\cr},$$
$$
\hat G_C=\pmatrix{(1-s_1s_3)^{-1}e^{-t_0A'}&\cr&(1-s_1s_3)e^{t_0A'}\cr}.$$

Taking into account the conditions (\ref{generic}), we may choose the 
reserved parameters $c$ and $t_0$ in such a way that 
\begin{equation}\label{p423}
s_3=e^{t_0B'/2},\quad
1-s_1s_3=e^{-t_0A'}.
\end{equation}
The system (\ref{p423}) yields immediately the parameter values
\begin{eqnarray}\label{p424}
&&\hskip-20pt
t_0=-2\frac{\omega_a}{\Delta}\ln s_3-
\frac{\omega_b}{\Delta}\ln (1-s_1s_3)
\pmod{\frac{4\pi i\omega_a}{\Delta}}
\pmod{\frac{2\pi i\omega_b}{\Delta}},\nonumber
\\
&&\hskip-20pt
\frac{c}{\omega_a}t_0=
2\frac{A}{\Delta}\ln s_3+\frac{B}{\Delta}\ln (1-s_1s_3)
\pmod{\frac{4\pi iA}{\Delta}}
\pmod{\frac{2\pi iB}{\Delta}},
\end{eqnarray}
where
$$
\Delta=A\omega_b-B\omega_a=
4i\bigl({\cal J}_a\omega_b-{\cal J}_b\omega_a\bigr)=
-\frac{16\pi}{3}\bigl(z_1^2+z_3^2\bigr)=\frac{8\pi}{3}e^{i\varphi}$$
due to the Riemann bilinear identity. 

Under the conditions (\ref{p423}), the RH problem for the 
function $\Pso^o$ simplifies dramatically:

1) near infinity, it behaves like the exponent
\begin{equation}\label{p425}
\Pso^o(z)=\bigl(I+{\cal O}(z^{-1})\bigr)e^{-(t-t_0)\theta\sigma_3},\quad 
z\to\infty;
\end{equation}

2) it jumps on the intervals $(-z_1,-z_3)$ and $(z_3,z_1)$:
\begin{eqnarray}\label{p426}
z\in (-z_1,-z_3)\colon&&
\Pso^o{\!}_+(z)=\Pso^o{\!}_-(z)\pmatrix{&-1\cr1&\cr},
\\\label{p427}
z\in (z_3,z_1)\colon&&
\Pso^o{\!}_+(z)=\Pso^o{\!}_-(z)\pmatrix{&1\cr-1&\cr};
\end{eqnarray}

3) it has no more than $L_{2}$-integrable
singularities at $\pm z_1$, $\pm z_3$.

Apart from the countable set of values of $t$ described below, this problem 
can be solved explicitly, cf.\ \cite{DIZ}. Let us introduce the function 
\begin{equation}\label{p428}
\mu=\frac{1}{2}\pmatrix{\beta+\beta^{-1}&-i\bigl(\beta-\beta^{-1}\bigr)\cr
i\bigl(\beta-\beta^{-1}\bigr)&\beta+\beta^{-1}\cr},
\end{equation}
where
\begin{equation}\label{p429}
\beta=\Bigl(\frac{z^2-z_1^2}{z^2-z_3^2}\Bigr)^{1/4}
\end{equation}
is defined on the upper sheet of the Riemann surface $\Gamma$ (see Figure~10)
in such a way that
$$
\beta\to1\quad\hbox{as}\quad 
z\to\infty,$$
so that
\begin{equation}\label{p430}
\mu\longto_{z\to\infty}I.
\end{equation}
The discontinuity of the scalar function $\beta$ is described by the relations
\begin{eqnarray}
z\in(-z_1,-z_3)\colon&&
\beta_+=-i\beta_-,\nonumber
\\
z\in(z_3,z_1)\colon&&
\beta_+=i\beta_-,\nonumber
\end{eqnarray}
and therefore, for the function $\mu$, the jump conditions hold:
\begin{eqnarray}\label{p431}
z\in(-z_1,-z_3)\colon&&
\mu_+=\mu_-\pmatrix{&-1\cr1&\cr},\nonumber
\\
z\in(z_3,z_1)\colon&&
\mu_+=\mu_-\pmatrix{&1\cr-1&\cr},
\end{eqnarray}
The function $\Pso^o$ is constructed in terms of the Baker-Akhiezer
functions as follows:
\begin{equation}\label{p434}
\Pso^o(z)=
\pmatrix{\dsp\mu_{11}{\!}\Pso^{(1)}{\!}_{BA}& 
\dsp\mu_{12}{\!}\Pso^{(1)}{\!}_{BA}^{*}\cr
\dsp\mu_{21}{\!}\Pso^{(2)}{\!}_{BA}& 
\dsp\mu_{22}{\!}\Pso^{(2)}{\!}_{BA}^{*}\cr}
\end{equation}
where
\begin{eqnarray}\label{p435}
&&\Pso^{(1)}{\!}_{BA}(z)=
e^{-(t-t_0)\Omega(z)}
\frac{\Theta\bigl(U(z)+V(t-t_0)+\frac{1}{2}\bigr)
\Theta(\frac{1}{2})}
{\Theta\bigl(U(z)+\frac{1}{2}\bigr)
\Theta\bigl(V(t-t_0)+\frac{1}{2}\bigr)},\nonumber
\\
&&\Pso^{(2)}{\!}_{BA}=
e^{-(t-t_0)\bigl(\Omega(z)-\pi iV\bigr)}
\frac{\Theta\bigl(U(z)+V(t-t_0)+\frac{1+\tau}{2}\bigr)
\Theta\bigl(\frac{1}{2})}
{\Theta\bigl(U(z)+\frac{1+\tau}{2}\bigr)
\Theta\bigl(V(t-t_0)+\frac{1}{2}\bigr)},\nonumber
\\
&&\Pso^{(1)}{\!}_{BA}^{*}(z)=
e^{(t-t_0)\bigl(\Omega(z)-\pi iV\bigr)}
\frac{\Theta\bigl(U(z)-V(t-t_0)+\frac{1+\tau}{2}\bigr)
\Theta(\frac{1}{2})}
{\Theta\bigl(U(z)+\frac{1+\tau}{2}\bigr)
\Theta\bigl(V(t-t_0)+\frac{1}{2}\bigr)},\nonumber
\\
&&\Pso^{(2)}{\!}_{BA}^{*}=
e^{(t-t_0)\Omega(z)}
\frac{\Theta\bigl(U(z)-V(t-t_0)+\frac{1}{2}\bigr)
\Theta\bigl(\frac{1}{2})}
{\Theta\bigl(U(z)+\frac{1}{2}\bigr)
\Theta\bigl(V(t-t_0)+\frac{1}{2}\bigr)},
\end{eqnarray}
and
\begin{eqnarray}\label{p436}
&&\Omega(z)=g(z)-AU(z),\quad
U(z)=\frac{1}{\omega_a}\int_{\infty^+}^{z}\frac{dz}{w(z)},\quad
\tau=\frac{\omega_b}{\omega_a},
\\
&&V=\frac{\Delta}{2\pi i\omega_a}=
\frac{8i}{3\omega_a}\bigl(z_1^2+z_3^2\bigr)=
-\frac{4i}{3\omega_a}e^{i\varphi},\nonumber
\\\label{poles}
&&t\neq t_0+\frac{\tau}{2V}\pmod{\frac{1}{V}}\pmod{\frac{\tau}{V}},
\end{eqnarray}
and $\Theta(z)$ means the Riemann theta-function of $\Gamma$, 
$\dsp\Theta(z)=\sum_{n\in{\Bbb Z}}e^{\pi in^2\tau+2\pi inz}$.

It is worth to note that all $\Pso^{(k)}{\!}_{BA}(z)$, 
$\Pso^{(k)}{\!}_{BA}^{*}(z)$ are single valued on ${\Bbb C}^*$ cut along
$[z_3,z_1]\cup[-z_1,-z_3]$, see Figure~10, because $\oint_ad\Omega=0$. Also, 
considered on the Riemann surface, 
$\Pso^{(k)}{\!}_{BA}(z)=\Pso^{(k)}{\!}_{BA}(P)$ and 
$\Pso^{(k)}{\!}_{BA}^{*}(z)=\Pso^{(k)}{\!}_{BA}(P^*)$.

Now, let us verify that the function (\ref{p434}) yield the solution of the
RH problem (\ref{p425})--(\ref{p426}). Note first of all that $\Pso^o(z)$ has 
no pole on the upper sheet of $\Gamma$. Indeed, 
$\Theta\bigl(U(z)+\frac{1}{2}\bigr)\neq0$ $\forall z\in{\Bbb C}^*$. Taking
into account the equation $\Theta\bigl(\frac{1+\tau}{2}\bigr)=0$, the only 
zero of $\Theta\bigl(U(z)+\frac{1+\tau}{2}\bigr)$ lies at $\infty^+$. Thus 
$\Theta\bigl(U(z)+\frac{1+\tau}{2}\bigr)={\cal O}(z^{-1})$ while both 
$\mu_{12}$ and $\mu_{21}$ vanish as ${\cal O}(z^{-2})$.

Next, to check the asymptotics of $\Pso^o(z)$ near infinity on the upper 
sheet of $\Gamma$, we find
$$
U(z)=-\frac{1}{\omega_az}+{\cal O}(z^{-2}),\quad 
z\to\infty^+,$$ 
and therefore
$$
\bigl(\Pso^o\bigr)_{11}=e^{-i(t-t_0)(\frac{4}{3}z^3+e^{i\varphi}z)}
\bigl(1+{\cal O}(z^{-1})\bigr),$$
$$
\bigl(\Pso^o\bigr)_{22}=e^{i(t -t_0)(\frac{4}{3}z^3+e^{i\varphi}z)}
\bigl(1+{\cal O}(z^{-1})\bigr).$$
Similarly, using the asymptotics 
$$
\mu_{12}=-i\frac{z_3^2-z_1^2}{4z^2}+{\cal O}(z^{-4}),\quad
\mu_{21}=i\frac{z_3^2-z_1^2}{4z^2}+{\cal O}(z^{-4}),$$
$$
\Theta\bigl(U(z)+\frac{1+\tau}{2}\bigr)=
-\frac{1}{\omega_az}\Theta'\bigl(\frac{1+\tau}{2}\bigr)+
{\cal O}(z^{-2}),$$
we obtain
$$
\bigl(\Pso^o\bigr)_{21}=e^{-i(t-t_0)(\frac{4}{3}z^3+e^{i\varphi}z)}\cdot
{\cal O}(z^{-1}),\quad
\bigl(\Pso^o\bigr)_{12}=e^{i(t-t_0)(\frac{4}{3}z^3+e^{i\varphi}z)}\cdot
{\cal O}(z^{-1}),$$
and the asymptotics (\ref{p425}) holds true. To check the jump conditions 
(\ref{p426}), (\ref{p427}), it is enough, due to (\ref{p431}), to show that
$$
\Pso^{(k)}{\!}_{BA}(z)_{\pm}=\Pso^{(k)}{\!}_{BA}^*(z)_{\mp}$$
on the segments $(z_3,z_1)$ and $(-z_1,-z_3)$. Let us consider the first of 
them. One can see that on the segment $(z_3,z_1)$,
$$
U_++U_-=\frac{\tau}{2},\quad
\Omega_++\Omega_-=\frac{1}{2}(B-A\tau)=
-\frac{\Delta}{2\omega_a}=-\pi iV.$$
Therefore
$$
\Pso^{(1)}{\!}_{BA}(z)_+=e^{-(t-t_0)\Omega_+(z)}
\frac{
\Theta\bigl(U_+(z)+V(t-t_0)+\frac{1}{2}\bigr)
\Theta(\frac{1}{2})}
{\Theta\bigl(U_+(z)+\frac{1}{2}\bigr)
\Theta\bigl(V(t-t_0)+\frac{1}{2}\bigr)}=$$
$$
=e^{(t-t_0)\bigl(\Omega_-(z)+\pi iV\bigr)}
\frac{
\Theta\bigl(-U_-(z)+\frac{\tau}{2}+V(t-t_0)+\frac{1}{2}\bigr)
\Theta(\frac{1}{2})}
{\Theta\bigl(-U_-(z)+\frac{\tau}{2}+\frac{1}{2}\bigr)
\Theta\bigl(V(t-t_0)+\frac{1}{2}\bigr)}=$$
$$
=e^{(t-t_0)\bigl(\Omega_-(z)-\pi iV\bigr)}
\frac{\Theta\bigl(U_-(z)+\frac{1+\tau}{2}-V(t-t_0)\bigr)
\Theta(\frac{1}{2})}
{\Theta\bigl(U_-(z)+\frac{1+\tau}{2}\bigr)
\Theta\bigl(V(t-t_0)+\frac{1}{2}\bigr)}=
\Pso^{(1)}{\!}_{BA}^*(z)_-,$$
where we have used the properties $\Theta(-z)=\Theta(z)$, 
$\Theta(z+1)=\Theta(z)$, 
$\Theta(z\pm\tau)=e^{-\pi i\tau\mp2\pi iz}\Theta(z)$. The other conditions 
can be checked similarly.

Thus we obtain the following representation for the function $\Psi^{(BA)}$:
\begin{eqnarray}\label{p437}
&&\Psi^{(BA)}=\Pso^o e^{-t_0h\sigma_3}=
\pmatrix{\dsp\mu_{11}{\!}\Pso^{(1)}{\!}_{BA}& 
\dsp\mu_{12}{\!}\Pso^{(1)}{\!}_{BA}^{*}\cr
\dsp\mu_{21}{\!}\Pso^{(2)}{\!}_{BA}& 
\dsp\mu_{22}{\!}\Pso^{(2)}{\!}_{BA}^{*}\cr}
e^{-t_0h\sigma_3}=
\\
&&=\pmatrix{\dsp\mu_{11}{\!}\Theto^{(1)}& 
\dsp\mu_{12}e^{-\pi iV(t -t_0)}\Theto^{(1)}{\!}^{*}\cr
\dsp\mu_{21}e^{\pi iV(t -t_0)}\Theto^{(2)}& 
\dsp\mu_{22}{\!}\Theto^{(2)}{\!}^{*}\cr}
e^{(t A-t_0A')U\sigma_3}e^{-t g\sigma_3}\,,\nonumber
\end{eqnarray}
where
$$
\Theto^{(1)}\!=\!
\frac{\Theta\bigl(U(z)+V(t-t_0)+\frac{1}{2}\bigr)
\Theta(\frac{1}{2})}
{\Theta\bigl(U(z)+\frac{1}{2}\bigr)
\Theta\bigl(V(t-t_0)+\frac{1}{2}\bigr)},\
\Theto^{(1)}{\!}^{*}\!=\!
\frac{\Theta\bigl(U(z)-V(t-t_0)+\frac{1+\tau}{2}\bigr)
\Theta(\frac{1}{2})}
{\Theta\bigl(U(z)+\frac{1+\tau}{2}\bigr)
\Theta\bigl(V(t-t_0)+\frac{1}{2}\bigr)},$$
$$
\Theto^{(2)}\!=\!
\frac{\Theta\bigl(U(z)+V(t-t_0)+\frac{1+\tau}{2}\bigr)
\Theta\bigl(\frac{1}{2})}
{\Theta\bigl(U(z)+\frac{1+\tau}{2}\bigr)
\Theta\bigl(V(t-t_0)+\frac{1}{2}\bigr)},\
\Theto^{(2)}{\!}^{*}\!=\!
\frac{\Theta\bigl(U(z)-V(t-t_0)+\frac{1}{2}\bigr)
\Theta\bigl(\frac{1}{2})}
{\Theta\bigl(U(z)+\frac{1}{2}\bigr)
\Theta\bigl(V(t-t_0)+\frac{1}{2}\bigr)}.$$

Note finally, that the determinant of $\Psi^{(BA)}(z)$ might have 
singularities at the branch points $\pm z_1$, $\pm z_3$ of order 
${\cal O}((z-z_k)^{-1/2})$. However, because of the asymptotics 
$\det\Psi^{(BA)}(z)=1+{\cal O}\bigl(z^{-1}\bigr)$ as $z\to\infty$ and absence 
of jump, the determinant is entire function of $z$ which, due to the 
Liouville theorem, is constant, $\det\Psi^{(BA)}(z)\equiv1$.

\section{Local RH problems near the branch points}
\label{local_solutions}
\setcounter{equation}{0}

\subsection{RH problem solvable by the Airy functions}

The function $\Psi^{(BA)}(z)$ has branch points at $z=z_k$, $k=1,2,3,4$, and
does not approach the solution $\Psi(z)$ of the problem
(\ref{p21u})--(\ref{p24u}). In this section, we construct such an
approximation using the model problem related to the Airy functions. Consider 
the Wronski matrices of the independent solutions $y_i(\zeta)$ of the Airy 
equation $y''=\zeta y$:
$$
\Phi (\zeta)=\pmatrix{y_1(\zeta)&y_2(\zeta)\cr 
y_1'(\zeta)&y_2'(\zeta)\cr}.$$
Using asymptotic formulae presented in ref.\ \cite{BE}, one can introduce
the matrix functions $\Phi_k(\zeta)$ with the standard asymptotic behavior as
$\zeta\to\infty$,
\begin{eqnarray}\label{airy}
&&\Phi_k(\zeta)=\zeta^{-\frac{\sigma_3}{4}}\pmatrix{1&1\cr 1&-1\cr}
\bigl(I+{\cal O}(\zeta^{-3/2})\bigr)e^{\frac{2}{3}\zeta^{3/2}\sigma_3},\quad
\zeta^{3/2}\to\infty,
\\
&&\zeta\in\omega_k=\Bigl\{\zeta\in{\Bbb C}\colon\
\arg\zeta\in\bigl(\frac{2\pi}{3}(k-\frac{3}{2})+\epsilon,\ 
\frac{2\pi}{3}(k+\frac{1}{2})-\epsilon\bigr)\Bigr\},\quad
\epsilon>0.\nonumber
\end{eqnarray}
These functions are related to each other by the Stokes multipliers
\begin{equation}\label{airy_stokes}
\Phi_{k+1}(\zeta)=\Phi_k(\zeta)G_k,\quad
G_{2k-1}=\pmatrix{1&-i\cr0&1\cr},\quad
G_{2k}=\pmatrix{1&0\cr-i&1\cr},
\end{equation}
This collection of the matrix functions allows us to construct an exact 
solution $\Phi(\zeta)$ of the following RH problem 
(see Figure~12):

\smallskip
1) near infinity
\begin{equation}\label{p513}
\Phi=\zeta^{-\frac{\sigma_3}{4}}\pmatrix{1&1\cr1&-1\cr}
\bigl(I+{\cal O}(\zeta^{-3/2})\bigr)e^{\frac{2}{3}\zeta^{3/2}\sigma_3},\quad
\zeta^{3/2}\to\infty,
\end{equation}
where $\zeta^{3/2}$ is defined on the plane cut along the ray 
$\arg\zeta=-\pi/3$;

2) on the anti-Stokes rays $\arg\zeta=\frac{2\pi}{3}k$, $k=0,1,2$, oriented 
to infinity, the jump conditions hold,
\begin{equation}\label{p514}
\Phi_+=\Phi_-G_k,\quad
\arg\zeta=\frac{2\pi}{3}k,\quad
k=0,1,2;
\end{equation}

3) on the ray $\arg\zeta=-\pi/3$ oriented to infinity, the jump condition 
holds,
\begin{equation}\label{p515}
\Phi_+=\Phi_-G_{\infty},\quad
G_{\infty}=i\sigma_1=\pmatrix{&i\cr i&\cr}.
\end{equation}
    
\begin{figure}[hbt]\label{pf9}
\begin{center}
\mbox{\psbox{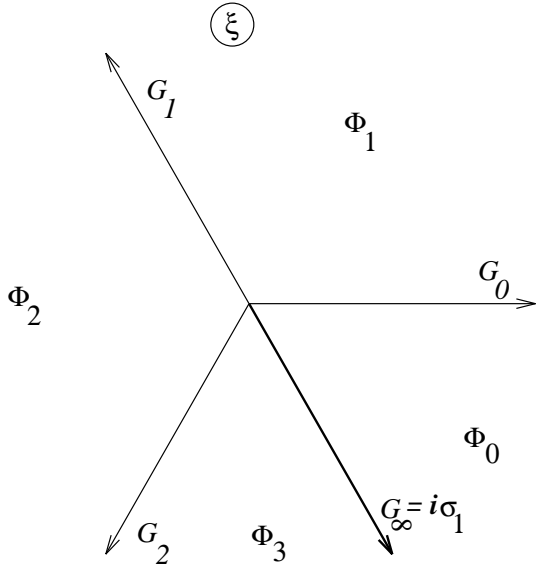}}
\caption{The RH problem graph for the collection of Wronski
matrices of the Airy functions.}
\end{center}
\end{figure}

\subsection{Parametrix in a neighborhood of the branch point $z_3$}

Let us change the variable in accord with
\begin{equation}\label{p519}
\zeta(z)=(-\frac{3t}{2}g_3(z))^{2/3},\quad
g_3(z)=4i\int_{z_3}^{z}w(z)\,dz\equiv 
g(z)-\frac{1}{4}B.
\end{equation}
This change is defined on the complex $z$-plane cut along the sum of 
intervals $[z_1,-z_1]\cup[-z_1,-z_3]\cup[-z_3,\infty)$ and mapping the 
domain ${\cal D}_3$ of $z$-plane bounded by the Stokes line chain shown in 
Figure~13 onto the complex $\zeta$-plane cut along the ray $\arg\zeta=-\pi/3$.
\begin{figure}[hbt]\label{pf10}
\begin{center}
\mbox{\psbox{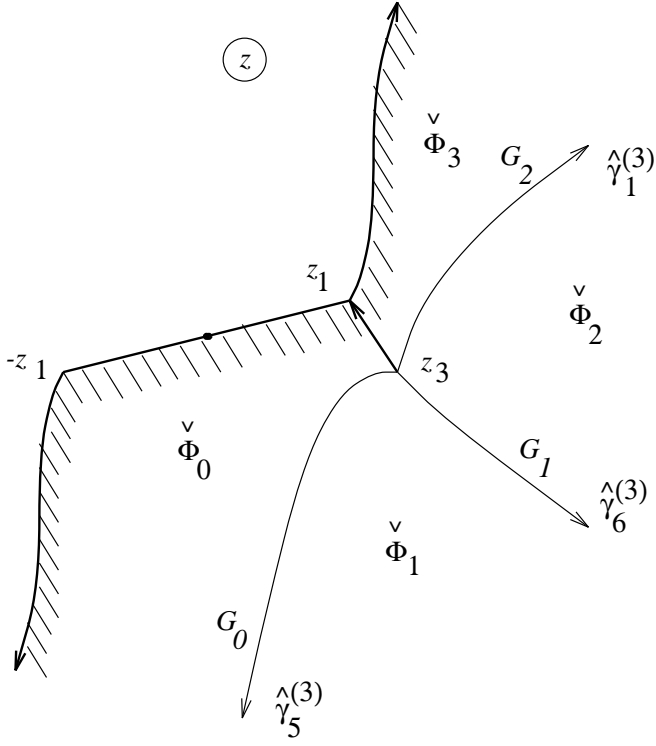}}
\caption{The jump graph for the fundamental system of the Airy functions 
after the change of variables $\zeta=\bigl(-3tg_3(z)/2\bigr)^{2/3}$.}
\end{center}
\end{figure}
Now, the function 
\begin{equation}\label{p521}
\check\Phi(z)=\Phi(\zeta(z))
\end{equation} 
which behaves at infinity as follows,
\begin{equation}\label{p522}
\check\Phi(z)=\bigl(-\frac{3t}{2}g_3(z)\bigr)^{-\frac{\sigma_3}{6}}
\pmatrix{1&1\cr 1&-1\cr}\bigl(I+{\cal O}(\frac{1}{tg_3(z)})\bigr)
e^{-tg_3(z)\sigma_3},\quad
tg_3(z)\to\infty,
\end{equation}
jumps on the graph shown in Figure~13.

The rays $\arg\zeta=2\pi k/3$, $k=0,1,2$, are transformed into the 
anti-Stokes lines $\hat\gamma_5^{(3)}$, $\hat\gamma_6^{(3)}$ and 
$\hat\gamma_1^{(3)}$, respectively, see Figure~4, moreover the jump matrices 
are preserved there. The cut $\arg\zeta=-\pi/3$ turns out the segment 
$(z_3,z_1)$ which splits at the point $z_1$ into the boundary of the 
$z$-domain.

Therefore, the function
\begin{equation}\label{p523}
\Phi^{(3)}(z)=\check\Phi(z)e^{i\frac{\pi}{4}\sigma_3}
s_3^{-\frac{\sigma_3}{2}},
\end{equation}
has the jumps on the graph shown on Figure~12 which is described by the 
matrices (\ref{p230u}), (\ref{p235u}),
\begin{eqnarray}\label{p511a}
z\in\hat\gamma_5^{(3)}\colon\
{\Eu S}_5^{(3)}=\pmatrix{1&0\cr\frac{1}{s_3}&1\cr};&&
z\in\hat\gamma_6^{(3)}\colon\
{\Eu S}_6=\pmatrix{1&-s_3\cr0&1\cr},\nonumber
\\
z\in\hat\gamma_1^{(3)}\colon\ 
{\Eu S}_1^{(3)}=\pmatrix{1&0\cr \frac{1}{s_3}&1\cr};&&
z\in(z_3,z_1)\colon\ 
{\Eu S}_B=\pmatrix{&{\scr s_3}\cr-\frac{1}{s_3}&\cr},
\end{eqnarray}
and has the asymptotics near infinity
\begin{eqnarray}\label{p524}
&&\Phi^{(3)}(z)=\bigl(-\frac{3t}{2}g_3(z)\bigr)^{-\frac{\sigma_3}{6}}
\pmatrix{1&1\cr 1&-1\cr}
\bigl(I+{\cal O}(\frac{1}{t g_3(z)})\bigr)\times\nonumber
\\
&&\times 
e^{\frac{1}{4}(tB-t_0B')\sigma_3}
e^{i\frac{\pi}{4}\sigma_3}e^{-tg\sigma_3},\quad
tg_3(z)\to\infty.
\end{eqnarray}
It is important, that the jumps of the matrix functions $\Psi(z)$ and
$\Phi^{(3)}(z)$ coincide with each other on the contours $\hat\gamma_5^{(3)}$,
$\hat\gamma_6^{(3)}$, $\hat\gamma_1^{(3)}$ and $\bigl(z_3,z_1\bigr)$, so that
they differ from each other in a left matrix multiplier $M^{(3)}(z)$ 
holomorphic in some bounded neighborhood of the node point $z=z_3$:
$$
\Psi(z)=M^{(3)}(z)\Phi^{(3)}(z).$$

Besides the Airy fundamental matrix set $\Phi(\xi)$, let us introduce its
reduced asymptotic form
$$
\Phi_{as}(\zeta)=\zeta^{-\frac{\sigma_3}{4}}\pmatrix{1&1\cr 1&-1\cr}
e^{\frac{2}{3}\zeta^{3/2}\sigma_3},$$
which solves the RH problem on the ray $\arg\zeta=-\pi/3$ 
oriented from zero to infinity,
\begin{equation}\label{phias}
\Phi_{as_+}=\Phi_{as_-}G_{\infty},\quad
G_{\infty}=i\sigma_1=\pmatrix{&i\cr i&\cr}.
\end{equation}
Using the changes (\ref{p519}), (\ref{p523}), the matrix
$$
\Phi_{as}^{(3)}(z)=\Phi_{as}(\zeta(z))e^{i\frac{\pi}{4}\sigma_3}
s_3^{-\frac{\sigma_3}{2}}$$
jumps on the Stokes segment $(z_3,z_1)$, and the jump matrix is (\ref{p235u})
$$
\Phi_{as_+}^{(3)}=\Phi_{as_-}^{(3)}{\Eu S}_B,\quad
{\Eu S}_B=\pmatrix{&{\scr s_3}\cr-\frac{1}{s_3}&\cr},\quad
z\in(z_3,z_1).$$
Since the jumps of the matrix functions $\Psi^{(BA)}(z)$ and
$\Phi_{as}^{(3)}(z)$ coincide with each other on the segment $(z_3,z_1)$, the 
functions differ from each other in a left matrix multiplier $N^{(3)}(z)$ 
holomorphic in some bounded neighborhood of the point $z=z_3$,
\begin{equation}\label{N3}
\Psi^{(BA)}(z)=N^{(3)}(z)\Phi_{as}^{(3)}(z).
\end{equation}
Let us define
\begin{equation}\label{psi3}
\Psi^{(3)}(z)=N^{(3)}(z)\Phi^{(3)}(z),\quad
N^{(3)}(z)=\Psi^{(BA)}(z)\bigl(\Phi_{as}^{(3)}(z)\bigr)^{-1}.
\end{equation}

\subsection{Parametrix in a neighborhood of the branch point $z_1$}

To construct the parametrix near $z=z_1$, let us introduce the function 
$\hat\Phi^{(1)}(\zeta)$,
\begin{equation}\label{hat_phi1}
\hat\Phi^{(1)}(\zeta)=\cases{
\Phi_0(\zeta)e^{-i\frac{\pi}{4}\sigma_3}
\bigl(\frac{s_3}{1-s_1s_3}\bigr)^{-\frac{\sigma_3}{2}}
\quad\hbox{ for }\quad\arg\zeta\in\bigl(-\frac{\pi}{3},0\bigr),\cr
\Phi_1(\zeta)e^{-i\frac{\pi}{4}\sigma_3}
\bigl(\frac{s_3}{1-s_1s_3}\bigr)^{-\frac{\sigma_3}{2}}
\quad\hbox{ for }\quad\arg\zeta\in\bigl(0,\frac{2\pi}{3}\bigr),\cr
\Phi_2(\zeta)e^{-i\frac{\pi}{4}\sigma_3}
\bigl(\frac{s_3}{1-s_1s_3}\bigr)^{-\frac{\sigma_3}{2}}
\quad\hbox{ for }\quad\arg\zeta\in\bigl(\frac{2\pi}{3},\pi\bigr),\cr
\Phi_2(\zeta)e^{-i\frac{\pi}{4}\sigma_3}
\bigl(s_3(1-s_1s_3)\bigr)^{-\frac{\sigma_3}{2}}
\quad\hbox{ for }\quad\arg\zeta\in\bigl(\pi,\frac{4\pi}{3}\bigr),\cr
\Phi_3(\zeta)e^{-i\frac{\pi}{4}\sigma_3}
\bigl(s_3(1-s_1s_3)\bigr)^{-\frac{\sigma_3}{2}}
\quad\hbox{ for }\quad
\arg\zeta\in\bigl(\frac{4\pi}{3},\frac{5\pi}{3}\bigr).\cr}
\end{equation}
As it is easy to see, the introduced matrix function satisfies the following 
jump conditions on the rays oriented to infinity,
\begin{eqnarray}\label{p514(1)}
&&\hat\Phi_+^{(1)}(\zeta)=\hat\Phi_-^{(1)}(\zeta){\Eu S}_1^{(1)},\quad
{\Eu S}_1^{(1)}=\pmatrix{1&0\cr-\frac{1-s_1s_3}{s_3}&1\cr},\quad
\arg\zeta=0,
\\
&&\hat\Phi_+^{(1)}(\zeta)=\hat\Phi_-^{(1)}(\zeta){\Eu S}_2^{(1)},\quad
{\Eu S}_2^{(1)}=\pmatrix{1&\frac{s_3}{1-s_1s_3}\cr0&1\cr},\quad
\arg\zeta=\frac{2\pi}{3},\nonumber
\\
&&\hat\Phi_+^{(1)}(\zeta)=\hat\Phi_-^{(1)}(\zeta){\Eu S}_C,\quad
{\Eu S}_C=\pmatrix{\frac{1}{1-s_1s_3}&\cr &1-s_1s_3\cr},\quad
\arg\zeta=\pi,\nonumber
\\
&&\hat\Phi_+^{(1)}(\zeta)=\hat\Phi_-^{(1)}(\zeta){\Eu S}_5^{(1)},\quad
{\Eu S}_5^{(1)}=\pmatrix{1&0\cr -\frac{1}{s_3(1-s_1s_3)}&1\cr},\quad
\arg\zeta=\frac{4\pi}{3},\nonumber
\\
&&\hat\Phi_+^{(1)}(\zeta)=\hat\Phi_-^{(1)}(\zeta){\Eu S}_B^{-1},\quad
{\Eu S}_B^{-1}=\pmatrix{&-s_3\cr \frac{1}{s_3}&\cr},\quad
\arg\zeta=-\frac{\pi}{3}.\nonumber
\end{eqnarray}

Introduce the change of variables
\begin{equation}\label{p525}
\zeta(z)=\bigl(-\frac{3t}{2}g_1(z)\bigr)^{2/3},
\end{equation}
where
$$
g_1(z)=4i\int_{z_1}^{z}w(z)\,dz$$
is defined on $z$-plane cut along
$(\infty,z_3)\cup(z_3,z_1)\cup(\infty,-z_1)\cup(-z_1,-z_3)$. This change is
holomorphic near the point $z_1$, and hence it is defined on the $z$-plane cut
along the sum of intervals $(z_3,\infty)\cup(\infty,-z_1)\cup(-z_1,-z_3)$ and
maps the domain ${\cal D}_1$ of $z$-plane with the boundary on the Stokes 
lines shown on Figure~14, onto the complex $\zeta$-plane cut along the rays 
$\arg\zeta=-\pi/3$, $\arg(\zeta-\zeta_2)=\pi$, with some $\zeta_2$, 
$\arg\zeta_2=\pi$. 
\begin{figure}[hbt]\label{pf11}
\begin{center}
\mbox{\psbox{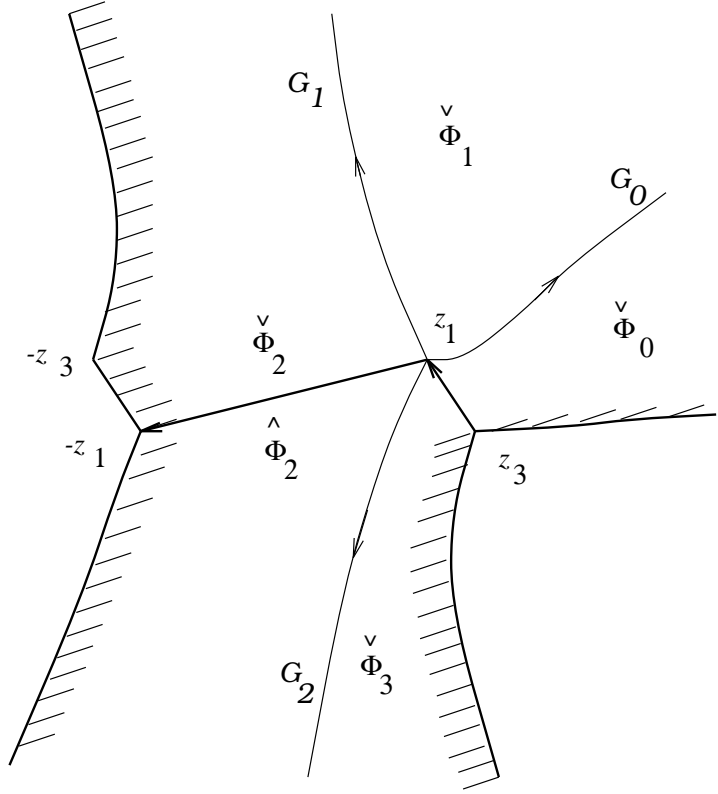}}
\caption{The jump graph for the set of the Wronski matrices of the Airy
functions after the change of variables 
$\zeta=\bigl(-3tg_1(z)/2\bigr)^{2/3}$.}
\end{center}
\end{figure}
Thus, in the area to the right of the cut $(z_3,z_1)\cup(z_1,-z_1)$, see 
Figure~13, or, equivalently, for 
$\arg\bigl(-\frac{3t}{2}g_1(z)\bigr)\in(-\frac{\pi}{2},\frac{3\pi}{2})$,
\begin{equation}\label{gg1-}
g_1(z)=g(z)-\frac{1}{4}B-\frac{1}{2}A,
\end{equation}
and in the area to the left of the cut $(z_3,z_1)\cup(z_1,-z_1)$ i.e.\ for 
$z$ satisfying the condition
$\arg\bigl(-\frac{3t}{2}g_1(z)\bigr)\in(\frac{3\pi}{2},\frac{5\pi}{2})$,
\begin{equation}\label{gg1+}
g_1(z)=g(z)-\frac{1}{4}B+\frac{1}{2}A.
\end{equation}

The function 
\begin{equation}\label{p527}
\Phi^{(1)}(z)=\hat\Phi^{(1)}(\zeta(z))
\end{equation} 
with the asymptotics at infinity
\begin{eqnarray}\label{p528}
&&\Phi^{(1)}(z)=\bigl(-\frac{3t}{2}g_1(z)\bigr)^{-\frac{\sigma_3}{6}}
\pmatrix{1&1\cr1&-1\cr}
\bigl(I+{\cal O}(\frac{1}{t g_1(z)})\bigr)\times\nonumber
\\
&&\times
e^{-tg\sigma_3}e^{-i\frac{\pi}{4}\sigma_3}
e^{\frac{1}{4}(tB-t_0 B')\sigma_3}
e^{\frac{1}{2}(tA-t_0 A')\sigma_3},\nonumber
\\
&&tg_1(z)\to\infty,\quad
\arg\bigl(-tg_1(z)\bigr)\in\bigl(-\frac{\pi}{2},\frac{3\pi}{2}\bigr),\nonumber
\\
&&\Phi^{(1)}(z)=\bigl(-\frac{3t}{2}g_1(z)\bigr)^{-\frac{\sigma_3}{6}}
\pmatrix{1&1\cr1&-1\cr}
\bigl(I+{\cal O}(\frac{1}{t g_1(z)})\bigr)\times\nonumber
\\
&&\times
e^{-t g\sigma_3}e^{-i\frac{\pi}{4}\sigma_3}
e^{\frac{1}{4}(t B-t_0 B')\sigma_3}
e^{-\frac{1}{2}(tA-t_0 A')\sigma_3},
\\
&&tg_1(z)\to\infty,\quad
\arg\bigl(-tg_1(z)\bigr)\in\bigl(\frac{3\pi}{2},\frac{5\pi}{2}\bigr),\nonumber
\end{eqnarray}
jumps on the graph shown on Figure~14. Under the change (\ref{p525}), the
interval $(z_3,z_1)$ corresponds to a segment of the ray 
$\arg\zeta=-\frac{\pi}{3}$ (if the orientation of the ray is from infinity to
the origin), the curve $\hat\gamma_1^{(1)}$ turns into the ray $\arg\zeta=0$,
the curve $\hat\gamma_2^{(1)}$ transforms into the ray
$\arg\zeta=\frac{2\pi}{3}$, the Stokes line $(z_1,-z_1)$ corresponds to the
ray $\arg\zeta=\pi$ and the curve $\hat\gamma_5^{(1)}$ corresponds to the ray
$\arg\zeta=\frac{4\pi}{3}$.

Since the discontinuities of the functions $\Psi(z)$ and $\Phi^{(1)}(z)$ are
the same on the contours $\hat\gamma_k^{(1)}$ and the Stokes segments
$(z_3,z_1)$ and $(z_1,-z_1)$, two these functions coincide with each other
up to a left matrix multiplier $M^{(1)}(z)$ holomorphic in some bounded
neighborhood of the point $z_1$:
$$
\Psi(z)=M^{(1)}(z)\Phi^{(1)}(z).$$

Besides the model function $\Phi^{(1)}(z)$, we introduce its reduced 
asymptotic form $\Phi_{as}^{(1)}(z)$ as follows:
\begin{eqnarray}\label{p528as}
&&\Phi_{as}^{(1)}(z)=
\bigl(-\frac{3t}{2}g_1(z)\bigr)^{-\frac{\sigma_3}{6}}
\pmatrix{1&1\cr1&-1\cr}\times\nonumber
\\
&&\times
e^{-tg\sigma_3}e^{-i\frac{\pi}{4}\sigma_3}
e^{\frac{1}{4}(tB-t_0 B')\sigma_3}
e^{\frac{1}{2}(tA-t_0 A')\sigma_3},\nonumber
\\
&&tg_1(z)\to\infty,\quad
\arg\bigl(-tg_1(z)\bigr)\in\bigl(-\frac{\pi}{2},\frac{3\pi}{2}\bigr),\nonumber
\\
&&\Phi_{as}^{(1)}(z)=
\bigl(-\frac{3t}{2}g_1(z)\bigr)^{-\frac{\sigma_3}{6}}
\pmatrix{1&1\cr1&-1\cr}\times\nonumber
\\
&&\times
e^{-tg\sigma_3}e^{-i\frac{\pi}{4}\sigma_3}
e^{\frac{1}{4}(tB-t_0 B')\sigma_3}
e^{-\frac{1}{2}(tA-t_0 A')\sigma_3},
\\
&&tg_1(z)\to\infty,\quad
\arg\bigl(-g_1(z)\bigr)\in\bigl(\frac{3\pi}{2},\frac{5\pi}{2}\bigr),\nonumber
\end{eqnarray}
This reduced function is continuous across the anti-Stokes curves 
$\hat\gamma_k^{(1)}$, and is characterized by the jump conditions
\begin{eqnarray}\label{p514(1)as}
&&\Phi_{as_+}^{(1)}(z)=\Phi_{as_-}^{(1)}(z){\Eu S}_C,\quad
{\Eu S}_C=\pmatrix{\frac{1}{1-s_1s_3}&\cr &1-s_1s_3\cr},\quad
z\in\bigl(z_1,-z_1\bigr),\nonumber
\\
&&\Phi_{as_+}^{(1)}(z)=\Phi_{as_-}^{(1)}(z){\Eu S}_B\,,\quad
{\Eu S}_B=\pmatrix{&s_3\cr -\frac{1}{s_3}&\cr},\quad
z\in\bigl(z_3,z_1\bigr)\,.\nonumber
\end{eqnarray}
Note again, that the jumps of the functions $\Psi^{(BA)}(z)$ and
$\Phi_{as}^{(1)}(z)$ coincide on the segments $(z_3,z_1)$ and $(z_1,-z_1)$
and hence the functions coincide with each other up to a left multiplier
$N^{(1)}(z)$ holomorphic in some bounded neighborhood of the point $z_1$:
\begin{equation}\label{N1}
\Psi^{(BA)}(z)=N^{(1)}(z)\Phi_{as}^{(1)}(z)\,.
\end{equation}
Define
\begin{equation}\label{psi1}
\Psi^{(1)}(z)=N^{(1)}(z)\Phi^{(1)}(z)\,,\quad
N^{(1)}(z)=\Psi^{(BA)}(z)\Bigl(\Phi_{as}^{(1)}(z)\Bigr)^{-1}\,.
\end{equation}

The functions solving the model RH problem associated with the
branch points $-z_1$ and $-z_3$ can be constructed in the very same way,
however they can be easily defined by the use of the symmetry relations:
\begin{equation}\label{p533}
\Psi^{(2)}(z)=\sigma_2\Psi^{(1)}(-z)\sigma_2\,,\quad
\Psi^{(4)}(z)=\sigma_2\Psi^{(3)}(-z)\sigma_2\,.
\end{equation}

\section{Asymptotic solution of the main RH problem}
\label{app_RH_solution}
\setcounter{equation}{0}

In this section, we will construct the approximate solution $\tilde\Psi(z)$
of the RH problem (\ref{p21u})--(\ref{p235u}) and prove the solvability of
the latter.
To this end, let us introduce four not overlapping disks $B_k$, 
$k=1,2,3,4$, centered at the branch points $z_k$, circles 
$C_k=\partial B_k$, and define the piecewise holomorphic function 
$\tilde\Psi(z)$,
\begin{equation}\label{p71}
\tilde\Psi(z)=\cases{\Psi^{(k)}(z),\quad
z\in B_k\,,\quad k=1,2,3,4,\cr
\tilde\Psi(z)=\Psi^{(BA)}(z),\quad
z\in{\Bbb C}\backslash\cup_{k=1}^4B_k.\cr}
\end{equation}
The function $\tilde\Psi$ solves the RH problem on the contour
depicted on Figure~15:

\begin{figure}[hbt]\label{pf13}
\begin{center}
\mbox{\psbox{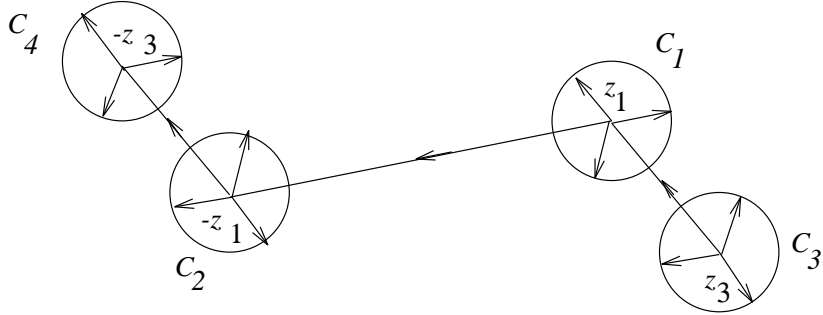}}
\caption{The model Riemann-Hilbert graph.}
\end{center}
\end{figure}

1) near infinity, due to (\ref{p411}),
\begin{equation}\label{p73}
\tilde\Psi(z)=\Psi^{(BA)}(z)=\bigl(I+{\cal O}(z^{-1})\bigr)
e^{-t\theta\sigma_3},\quad 
z\to\infty,
\end{equation}
where $\theta=i\bigl(\frac{4}{3}z^3+e^{i\varphi}z\bigr)$;

2) on the parts $\hat\gamma_s^{(k)}\cap B_k$ of anti-Stokes lines
$\hat\gamma_s^{(k)}$ enclosed in the interior $B_k$ of the circle $C_k$,
and on the segments $(z_3,z_1)$, $(z_1,-z_1)$, $(-z_1,-z_3)$, the jump
conditions hold
\begin{equation}\label{p74}
\tilde\Psi_+(z)=\tilde\Psi_-(z){\Eu S}
\end{equation}
with the jump matrices ${\Eu S}$ defined in (\ref{p230u})--(\ref{p235u});

3) on the circles $C_k$, $k=1,2,3,4$, oriented counter-clockwise, the jump 
conditions hold
\begin{equation}\label{p75}
\tilde\Psi_+(z)=\tilde\Psi_-(z)H(z),
\end{equation}
where, in accord with (\ref{N3}), (\ref{psi3}), (\ref{N1}), (\ref{psi1}),
(\ref{p210a}) and (\ref{p533}),
$$
H(z)=(\Psi^{(BA)}(z))^{-1}\Psi^{(k)}(z)=
(\Phi_{as}^{(k)}(z))^{-1}\Phi^{(k)}(z),\quad
z\in C_k.$$

To describe the difference between the model problem solution $\tilde\Psi(z)$
and the solution $\Psi(z)$ of the original problem, let us construct the 
``ratio"
\begin{equation}\label{p76}
Z(z)=\Psi(z)\tilde\Psi^{-1}(z).
\end{equation}

The correction function $Z(z)$ satisfies the RH problem on the jump contour
shown on Figure~16:

\begin{figure}[hbt]\label{pf14}
\begin{center}
\mbox{\psbox{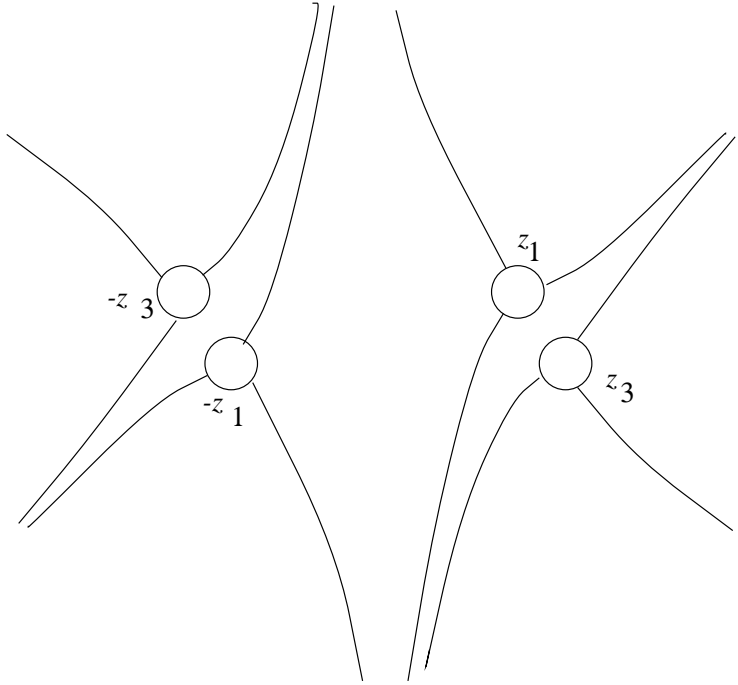}}
\caption{The RH graph for the correction function $Z(z)$.}
\end{center}
\end{figure}

\smallskip
\noindent
1) $Z(z)$ is normalized at infinity
\begin{equation}\label{p77}
Z\longto_{z\to\infty}I
\end{equation}

\noindent
2) on the parts of the anti-Stokes lines outside the circles $C_k$ and on the 
circles themselves, the following jump conditions hold true:
\begin{equation}\label{p78}
Z_+(z)=Z_-(z){\Eu G}_s^{(k)}(z),\quad
z\in\hat\gamma_s^{(k)}\cap\bigl({\Bbb C}\backslash B_k\bigr),
\end{equation}
where, on the anti-Stokes tails,
$$
{\Eu G}_s^{(k)}(z)=
\Psi^{(BA)}(z){\Eu S}_s^{(k)}\bigl(\Psi^{(BA)}(z)\bigr)^{-1},$$
and, on the circles $C_k$, $k=1,2,3,4$,
\begin{equation}\label{p79}
Z_+(z)=Z_-(z){\Eu G}_k(z),\quad
z\in C_k,
\end{equation}
where
$$
{\Eu G}_k(z)=\Psi^{(BA)}(z)\bigl(\Psi^{(k)}(z)\bigr)^{-1}=$$
$$
\qquad\qquad
=N^{(k)}(z)\Phi_{as}^{(k)}(z)\bigl(\Phi^{(k)}(z)\bigr)^{-1}
\bigl(N^{(k)}(z)\bigr)^{-1}.$$

\smallskip
As it is easy to see, for $t$ lying apart from the points (\ref{poles}), the 
jump matrices on the anti-Stokes lines have the asymptotics
$$
{\Eu G}_s^{(k)}(z)=I+\bigl(e^{-2|\scriptRe(t g(z))|}\bigr)=
I+\bigl(e^{-2|\scriptRe(t g_k(z))|}\bigr),$$
where we have used the Boutroux equations (\ref{p37}). For $t$ lying apart 
from the points (\ref{poles}), using the definition of $\Phi_{as}^{(k)}(z)$ 
and the boundedness of the matrices $N{(k)}(z)$ and their inverse, we find 
the estimate
$$
{\Eu G}_k(z)=I+(t^{-1}g_k^{-1}(z)).$$

Now, we are going to prove the solvability of the RH problem
(\ref{p77})--(\ref{p79}). 

First note that the jump matrices are continuous on the RH problem graph.
Consider the connected component of the graph associated with the point
$z_3$ (see Figure~17).

\begin{figure}[hbt]\label{pf15}
\begin{center}
\mbox{\psbox{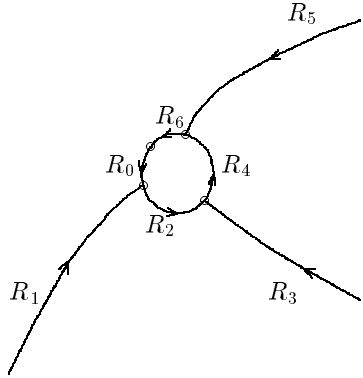}}
\caption{The connected component of the RH problem graph associated to the
point $z_3$.}
\end{center}
\end{figure}

Seeking for convenience, we orient the anti-Stokes lines $\hat\gamma_s^{(3)}$
in the direction from infinity to $z_3$. Denote provisionally the jump
matrices for the function $Z(z)$ as $R(z)$, i.e.\
$Z_+(z)=Z_-(z)R(z)$, moreover:

\smallskip
\noindent
on the part $C_{25}^{(3)}$ of the circle $C_3$ between the points of its
intersection with the Stokes line $(z_3,z_1)$ and anti-Stokes line
$\hat\gamma_5^{(3)}$, the jump matrix $R(z)=R_0(z)$ is as follows:
\begin{equation}\label{R_0}
R_0(z)=N^{(3)}(z)\Phi_{as}^{(3)}(z)
(\Phi_0^{(3)}(z))^{-1}(N^{(3)}(z))^{-1}=I+{\cal O}(t^{-1}g_3^{-1}(z));
\end{equation}

\noindent
on the anti-Stokes line $\hat\gamma_5^{(3)}$, the jump matrix $R(z)=R_1(z)$
is given by
\begin{equation}\label{R_1}
R_1(z)=\Psi^{(BA)}(z)({\Eu S}_5^{(3)})^{-1}(\Psi^{(BA)}(z))^{-1}=
I+{\cal O}(e^{2t g(z)});
\end{equation}

\noindent
on the part $C_{56}^{(3)}$ of the circle $C_3$ between the points of its
intersection with the anti-Stokes lines $\hat\gamma_5^{(3)}$ and
$\hat\gamma_6^{(3)}$, the jump matrix $R(z)=R_2(z)$ is as follows:
\begin{eqnarray}\label{R_2}
&&R_2(z)=N^{(3)}(z)\Phi_{as}^{(3)}(z)
(\Phi_1^{(3)}(z))^{-1}(N^{(3)}(z))^{-1}=\nonumber
\\
&&=N^{(3)}(z)\Phi_{as}^{(3)}(z)({\Eu S}_5^{(3)})^{-1}
(\Phi_0^{(3)}(z))^{-1}(N^{(3)}(z))^{-1}=\nonumber
\\
&&=R_1(z)R_0(z)=
I+{\cal O}(t^{-1}g_3^{-1}(z));
\end{eqnarray}

\noindent
on the anti-Stokes line $\hat\gamma_6^{(3)}$, the jump matrix $R(z)=R_3(z)$
is given by
\begin{equation}\label{R_3}
R_3(z)=\Psi^{(BA)}(z)({\Eu S}_6)^{-1}(\Psi^{(BA)}(z))^{-1}=
I+{\cal O}(e^{-2t g(z)});
\end{equation}

\noindent
on the part $C_{61}^{(3)}$ of the circle $C_3$ between the points of its
intersection with the anti-Stokes lines $\hat\gamma_6^{(3)}$ and
$\hat\gamma_1^{(3)}$, the jump matrix $R(z)=R_4(z)$ is as follows:
\begin{eqnarray}\label{R_4}
&&R_4(z)=N^{(3)}(z)\Phi_{as}^{(3)}(z)
(\Phi_2^{(3)}(z))^{-1}(N^{(3)}(z))^{-1}=\nonumber
\\
&&=N^{(3)}(z)\Phi_{as}^{(3)}(z)({\Eu S}_6)^{-1}({\Eu S}_5^{(3)})^{-1}
(\Phi_0^{(3)}(z))^{-1}(N^{(3)}(z))^{-1}=\nonumber
\\
&&=R_3(z)R_1(z)R_0(z)=I+{\cal O}(t^{-1}g_3^{-1}(z));
\end{eqnarray}

\noindent
on the anti-Stokes line $\hat\gamma_1^{(3)}$, the jump matrix $R(z)=R_5(z)$
is given by
\begin{equation}\label{R_5}
R_5(z)=\Psi^{(BA)}(z)({\Eu S}_1^{(3)})^{-1}(\Psi^{(BA)}(z))^{-1}=
I+{\cal O}(e^{-2t g(z)});
\end{equation}

\noindent
on the part $C_{12}^{(3)}$ of the circle $C_3$ between the points of its
intersection with the anti-Stokes line $\hat\gamma_1^{(3)}$ and the Stokes
line $(z_3,z_1)$, the jump matrix $R(z)=R_6(z)$ is as follows:
\begin{eqnarray}\label{R_6}
&&R_6(z)=N^{(3)}(z)\Phi_{as}^{(3)}(z)
(\Phi_3^{(3)}(z))^{-1}(N^{(3)}(z))^{-1}=\nonumber
\\
&&=N^{(3)}(z)\Phi_{as}^{(3)}(z)
({\Eu S}_1^{(3)})^{-1}({\Eu S}_6)^{-1}({\Eu S}_5^{(3)})^{-1}
(\Phi_0^{(3)}(z))^{-1}(N^{(3)}(z))^{-1}=\nonumber
\\
&&=R_5(z)R_3(z)R_1(z)R_0(z)=I+{\cal O}(t^{-1}g_3^{-1}(z)).
\end{eqnarray}
The relations 
\begin{equation}\label{factor}
R_2=R_1R_0,\quad
R_4=R_3R_2,\quad
R_6=R_5R_4
\end{equation}
above provide us with the continuity of the RH problem at the node points 
$C_3\cap\hat\gamma_s^{(3)}$ (see \cite{Z}). The fact is obvious because of
the possibility of analytical continuation of the jump matrices $R(z)$ from
the contours $C_3$, $\hat\gamma_s^{(3)}$ and the possibility to split the
graph at the node points in accord with (\ref{factor}) and to transform it 
into the union of the circle $C_3$ and of three smooth curves with the common 
endpoint $P=C_3\cap(z_3,z_1)$. As to the point $P$ itself, the continuity of 
the RH problem follows from the equation $R_0(z)=R_6(z)$. Indeed, at this 
point,
\begin{eqnarray}
&&R_0(z)=\Psi_+^{(BA)}(z)(\Phi_0^{(3)}(z))_+^{-1}(N^{(3)}(z))^{-1}=\nonumber
\\
&&\quad\quad
=\Psi_-^{(BA)}(z){\Eu S}_B
(\Phi_3^{(3)}(z){\Eu S}_B)_+^{-1}(N^{(3)}(z))^{-1}=\nonumber
\\
&&\quad\quad
=\Psi_-^{(BA)}(z)(\Phi_3^{(3)}(z))_+^{-1}(N^{(3)}(z))^{-1}=R_6(z),
\end{eqnarray}
where plus and minus mean the left and right limits on the segment 
$(z_3,z_1)$. In the very similar manner, one can check the continuity of the 
RH problem on the other connected components of the graph. 

All the jump matrices are uniformly close to identity. More exactly, for 
$t$ lying apart from the points (\ref{poles}), the estimates hold true,
\begin{equation}\label{Restimate}
\|R(z)-I\|<\cases{c_1e^{-2|\scriptRe(tg(z))|},\quad
z\in\hat\gamma_s^{(k)},\cr
c_1|t|^{-1}|g_k(z)|^{-1},\quad
z\in C_k,\cr}
\end{equation}
where the concrete value of the constant $c_1$ is not important for us. The 
estimates (\ref{Restimate}) imply that the differences $R(z)-I$ are square 
integrable on the RH problem graph 
$\Sigma=\cup_kC_k\cup_{k,s}\hat\gamma_s^{(k)}$, i.e.\ $R(z)-I\in L^2(\Sigma)$.

The RH problem (\ref{p77})--(\ref{p79}), i.e.\ 
\begin{eqnarray}\label{solv8}
&&Z\longto_{z\to\infty}I,\nonumber
\\
&&Z_+(z)=Z_-(z)R(z),\quad
z\in\Sigma,
\end{eqnarray}
is equivalent to the system of singular integral equations for the function 
$\rho(\zeta)\equiv Z_-(\zeta)$,
\begin{equation}\label{solv10}
\rho(z)=I+\frac{1}{2\pi i}\int_{\Sigma}
\rho(\zeta)(R(\zeta)-I)\frac{d\zeta}{\zeta-z_-},
\end{equation}
where $z_-$ denotes the right limit of $z$ on the contour of integration. In
the symbolic operator form, the system reads
\begin{equation}\label{solv11}
\rho=I+C_-\bigl[M\rho\bigr],
\end{equation}
where $C_-$ is the Cauchy operator, and $M$ is the operator of the right 
multiplication in the matrix $R(\zeta)-I\in L^2(\Sigma)$. Thus $M$ acts in
$L^2(\Sigma)$ and satisfies the estimate
\begin{equation}\label{solv12}
\|M\|_{L^2(\Sigma)}< c_2|t^{-1}|,
\end{equation}
where the precise value of the constant $c_2$ is not important for us. Because
the Cauchy operator is bounded in $L^2(\Sigma)$, see 
\cite{litvinchuk_spitkovski, beals_deift_tomei, Z}, we obtain the estimate
\begin{equation}\label{solv13}
\|C_-M\|_{L^2(\Sigma)}<c_3|t^{-1}|.
\end{equation}
Therefore, for any large enough $t$ lying apart from the points 
(\ref{poles}), the operator $C_-M$ is contracting, equation for the function 
$\chi=\rho-I$, 
\begin{equation}\label{solv14}
\chi=C_-[M]+C_-\bigl[M\chi\bigr],
\end{equation}
is solvable in the space $L^2(\Sigma)$ by iterations, and the estimate holds
true,
\begin{equation}\label{solv15}
\|\rho-I\|_{L^2(\Sigma)}<c|t|^{-1}.
\end{equation}
The solution of the RH problem (\ref{solv8}) for the function $Z(z)$ is
given by the Cauchy integral
$$
Z(z)=I+\frac{1}{2\pi i}\int_{\Sigma}
\rho(\zeta)(R(\zeta)-I)\frac{d\zeta}{\zeta-z},$$ 
which implies the asymptotic relation
\begin{equation}\label{solv16}
\Psi(z)=\bigl(I+{\cal O}((1+|z|)^{-1}t^{-1})\bigr)\tilde\Psi(z)
\end{equation}
in any closed subdomain of ${\Bbb C}\backslash\Sigma$.

\section{Asymptotics of the Painlev\'e function}
\label{as_P2_solution}
\setcounter{equation}{0}

Now we are ready to compute the asymptotics of the Painlev\'e function.
Using the expansion of $\Psi(\lambda)$ near infinity,
\begin{eqnarray}\label{p81}
&&\Psi(\lambda)=\bigl(I+
\frac{1}{\lambda}(-\frac{i}{2}D\sigma_3+\frac{1}{2}u\sigma_1)+
{\cal O}(\frac{1}{\lambda^2})\bigr)
e^{-i(\frac{4}{3}\lambda^3+x\lambda)\sigma_3}=\nonumber
\\
&&=\bigl(I+\frac{1}{t^{1/3}z}(
-\frac{i}{2}D\sigma_3+\frac{1}{2}u\sigma_1)+
{\cal O}(\frac{1}{t^{2/3}z^2})\bigr)
e^{-it(\frac{4}{3}z^3+e^{i\varphi}z)\sigma_3},
\end{eqnarray}
where $u$ is the Painlev\'e function
(cf. (\ref{nonlinintegral})), i.e.\ the solution of (\ref{p2}), 
$D=v^2-u^4-xu^2$ and 
\begin{equation}\label{p82}
\Psi(z)=\bigl(I+{\cal O}(\frac{1}{t z})\bigr)\tilde\Psi(z)=
\bigl(I+{\cal O}(t^{-1}z^{-1})\bigr)\Pso^o(z)e^{-t_0h(z)\sigma_3},
\end{equation}
to find the asymptotics of $u$ up to terms of order $t^{-1}$, it is enough to 
expand the Baker-Akhiezer function $\Pso^{(1)}{\!}_{BA}^*$ or 
$\Pso^{(2)}{\!}_{BA}$ near infinity:
\begin{eqnarray}\label{p83}
&&\hskip-4pt
\frac{u}{2t^{1/3}}=\lim_{z\to\infty^+}z 
e^{\pi i(t-t_0)V}
\frac{\Theta(U(z)+V(t-t_0)+\frac{1+\tau}{2})\Theta(\frac{1}{2})}
{\Theta(U(z)+\frac{1+\tau}{2})\Theta(V(t-t_0)+\frac{1}{2})}\mu_{21}+
{\cal O}(t^{-1})=\nonumber
\\
&&\hskip-4pt
=i\frac{\omega_a}{4}(z_1^2-z_3^2)e^{\pi i(t-t_0)V}
\frac{\Theta(V(t-t_0)+\frac{1+\tau}{2})\Theta(\frac{1}{2})}
{\Theta(V(t-t_0)+\frac{1}{2})\Theta'(\frac{1+\tau}{2})}+{\cal O}(t^{-1}).
\end{eqnarray}

Equations (\ref{p37}), (\ref{p424}), and (\ref{p83}) are transformed to the
statement of Theorem~\ref{theorem1} via routine manipulations with the 
elliptic functions and integrals.

\medskip
{\bf Acknowledgments.} The first co-author was supported in part by the
NSF, grant No.~DMS--9801608. The second co-author was supported in part by
the RFBR, grant No.~99--01--00687. The second co-author is grateful to the 
staff of the Mathematical Department of IUPUI for its hospitality during
his visit when this work was done.

\ifx\undefined\bysame
\newcommand{\bysame}{\leavevmode\hbox to3em{\hrulefill}\,}
\fi

\end{document}